%% file: IEEE-conference-template-062824.tex
\documentclass[10pt,conference]{IEEEtran}
\IEEEoverridecommandlockouts

\usepackage{cite}
\usepackage{amsmath,amssymb,amsfonts}
\usepackage{algorithm}
\usepackage{algorithmic}
\usepackage{graphicx}
\usepackage{textcomp}
\usepackage[table]{xcolor}
\usepackage[T1]{fontenc}
\usepackage[utf8]{inputenc}
\usepackage{booktabs}
\usepackage{multirow}
\usepackage{tcolorbox}
\usepackage{enumitem}
\usepackage{url}

\newtcolorbox{findingbox}{colback=black!4,colframe=black!55,boxrule=0.5pt,arc=1pt,left=5pt,right=5pt,top=4pt,bottom=4pt}
\def\BibTeX{{\rm B\kern-.05em{\sc i\kern-.025em b}\kern-.08em
    T\kern-.1667em\lower.7ex\hbox{E}\kern-.125emX}}

\begin{document}

\title{From General Agents to RCA Experts:\\
A Self-Evolving Harness for Root Cause Analysis}

\author{
\IEEEauthorblockN{Haiyu Huang\IEEEauthorrefmark{1}, Jiewei Lyu\IEEEauthorrefmark{2},
Zhihan Jiang\IEEEauthorrefmark{1}\IEEEauthorrefmark{3}, Jinyang Liu\IEEEauthorrefmark{3},\\
Xiao He\IEEEauthorrefmark{3}, Tieying Zhang\IEEEauthorrefmark{3},
Wu Xiang\IEEEauthorrefmark{3}, Michael Lyu\IEEEauthorrefmark{1}}
\IEEEauthorblockA{
\IEEEauthorrefmark{1}The Chinese University of Hong Kong,
\{hyhuang25, zhjiang22, lyu\}@cse.cuhk.edu.hk\\
\IEEEauthorrefmark{2}Individual Researcher, jiewei.lyu@gmail.com\\
\IEEEauthorrefmark{3}ByteDance,
\{jinyang.liu, xiao.hx, tieying.zhang, xiangwu\}@bytedance.com}
}

\maketitle  

\begin{abstract}
Automated root cause analysis (RCA) with large language models (LLMs) has drawn growing attention. Today, SREs typically automate RCA with LLMs in one of two ways: directly using a general-purpose agent (e.g., Codex or Claude Code) for diagnosis, or building a specialized RCA agent from scratch. As mainstream general agents grow more capable and iterate quickly, our quantitative study finds that the former now often surpasses the latter. Its accuracy, however, still falls short of production needs, and this gap stems mainly from the external adaptation layer outside the agent's general capabilities, namely the harness. We therefore argue that LLM-based RCA should focus on this external harness, reusing the strong general capabilities of a modern agent rather than rebuilding an agent from scratch. A key capability of such a harness is to self-evolve, accumulating system-specific experience from past diagnoses so that it gets better the more it is used. We introduce OpsHarness, a self-evolving RCA harness that turns diagnosis experience into reusable expertise. Its data plane combines layered operational knowledge with an idea-card tool library, while its control plane coordinates setup, diagnosis, evolution, and verification. During evolution, OpsHarness contrasts successful and failed trajectories, converts their evidence into atomic proposals, and admits updates only through a dual-gate verification process designed to prevent overfitting and regression. Across two public benchmarks and an industrial deployment, OpsHarness achieves 59.0\% top-1 accuracy, improving over a bare general agent by 63.4\% and over baseline RCA agents by 4.02$\times$.
\end{abstract}

\begin{IEEEkeywords}
root cause analysis, AIOps, LLM agents, agent harness, self-evolving
\end{IEEEkeywords}

\section{Introduction}

Modern software is built as large meshes of microservices. When one service degrades, the symptom (e.g., a latency spike or a surge of errors) often surfaces far from its cause and propagates across many dependencies. Root cause analysis (RCA) is the task of locating that cause from the metrics~\cite{Microscope}, logs~\cite{logad}, and traces~\cite{sigelman2010dapper} a system emits. It is one of the most time-pressured tasks in site reliability engineering. Even with mature observability tooling, a single production incident can take an on-call engineer tens of minutes to localize \cite{fightincidents22, faultdiagsurvey25}. 

Automated RCA has been studied for decades. Early methods localize faults with dependency graphs, causal inference, or spectrum analysis~\cite{microrca20, sage21, eadro23, baro24, nezha, cloudranger,microdiag}. With the rise of large language models (LLMs), more recent methods build specialized LLM-based RCA agents that plan, call tools, read telemetry, and reason about the cause~\cite{rcacopilot24, rcagent24, mabc24, flowofaction25,lasrca}. These methods typically start from a model and design the full pipeline from scratch, including the ReAct loop, code execution, and planning, to form a specialized RCA agent or framework.

However, LLM agent development has moved to a new stage. General-purpose agent frameworks such as Claude Code~\cite{claudecode25}, OpenAI Codex~\cite{openaicodex25}, and the open-source OpenCode~\cite{opencode25} now provide mature tool use, code execution, sandboxing, and long-horizon planning out of the box, and they iterate quickly. A mainstream best practice for an agent's general capabilities has therefore emerged, and re-designing these parts (i.e., building an agent from scratch) is hard to get better. We confirm this on RCA with a quantitative study (Section~\ref{subsec:harness-key}):
a bare flagship general-purpose agent is already often better than prior RCA agents that were carefully designed from scratch. 
On the same backbone (e.g., GPT~5.5), bare Codex answers 45\% of OpenRCA queries, well above the 32\% reached by the RCA-Agent proposed in OpenRCA \cite{openrca25}; measured across all tested benchmarks, this lead holds on every backbone we test.

\begin{figure}[t]
\centering
\includegraphics[width=\linewidth]{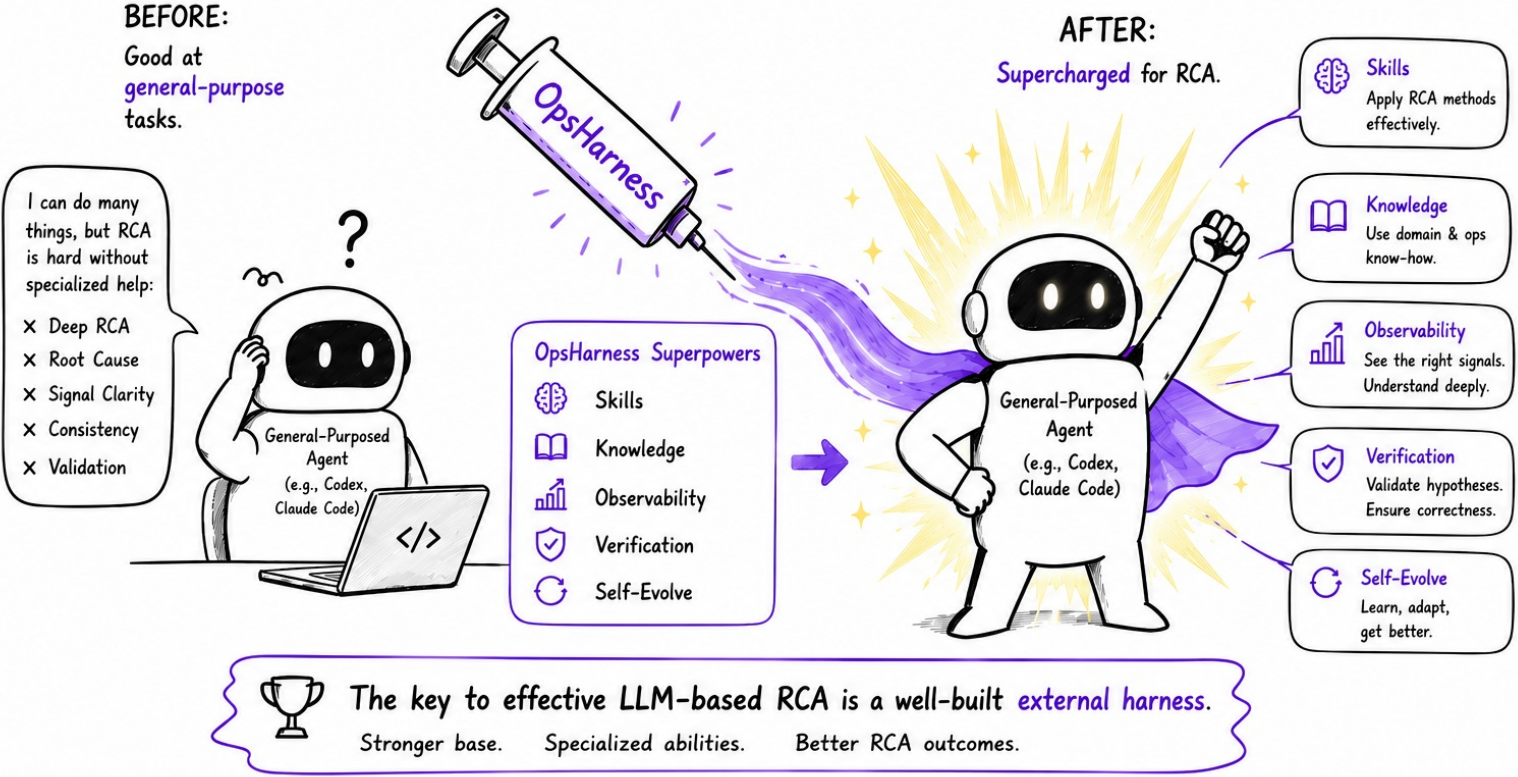}
\vspace{-0.20in}
\caption{The key to effective LLM-based RCA is a well-built external harness.}
\vspace{-0.20in}
\label{fig:intro}
\end{figure}

Nevertheless, even today's top-tier general-purpose agents still leave substantial room for improvement on RCA tasks, with Codex (GPT-5.5) achieving an average accuracy of only 51.2\% on the test datasets, as shown in our study in Sec.~\ref{subsec:harness-key}. Through our study of the failure cases, this room stems not from a deficit in general ability, but from the agent's lack of diagnostic expertise and unfamiliarity with the target system under diagnosis, much like a broadly capable engineer newly joining a team. Closing it is the responsibility not of the model or general ability of an agent, but of the \textbf{\emph{harness}}, the external software layer that wraps a general agent or model and adapts it to a specific task by supplying the tools, domain knowledge, and control logic it operates with~\cite{agentharness-anatomy26, sweagent24, contexteng25}. As models mature, a growing body of work finds that it is increasingly this surrounding layer, rather than the model or general ability alone, that decides how well an agent performs on a given task. The design of this surrounding layer has come to be known as \textbf{\emph{harness engineering}}~\cite{agentharness-anatomy26, contexteng25}.

Building on this observation, we argue that, in the current era of LLM agents, the key to effective LLM-based RCA lies in a well-built external harness (Fig.~\ref{fig:intro}), which should possess two properties. \textbf{First}, rather than rebuilding an RCA agent from scratch, the harness should focus on the diagnosis-specific infrastructure that sits outside the general capabilities, thereby
compatible with rapidly evolving general agents and reuses their strong general capabilities, while making them better at RCA. \textbf{Second, and more importantly}, this harness should continuously \textbf{\emph{self-evolve}}. As shown in our industrial study in Sec.~\ref{subsec:self-evolve}, what distinguishes an expert SRE from a capable newcomer is deep, accumulated experience of the target system, earned incident by incident (e.g., which metric pattern always means which fault here, how faults propagate between components)
Such experience is hard to specify in advance and usually has to be distilled from diagnosis over time. An effective harness should therefore be able to mine reusable best practices, knowledge, and tools from past diagnoses and apply them to later ones, so that, even in unfamiliar systems, it can gradually accumulate experience during diagnosis, self-evolve, and continuously improve.

To this end, we propose \textit{\textbf{OpsHarness}}, the first self-evolving external agent harness for RCA. Delivered as an external harness compatible with a general-purpose agent, it equips the agent with diagnosis-related skills, tools, and control loops to better handle RCA tasks. OpsHarness is organized into a \emph{data plane} and a \emph{control plane}. The data plane comprises a layered operational knowledge store 
and an idea-card tool library.
The control plane drives the harness lifecycle through four workflows, 
namely \emph{setup}, \emph{diagnose}, \emph{evolve} and \emph{verify}.
The evolve and verify workflows form a self-evolution loop. The evolve stage mines runtime diagnosis trajectories and distills patterns from both correct and incorrect runs into atomic evolution proposals. The verification stage then applies a dual-gate sandbox check to each proposal and only promotes those that pass, enabling OpsHarness to continuously improve while preventing overfitting.
We conduct comprehensive experiments on OpsHarness across two public benchmarks (i.e., OpenRCA~\cite{openrca25} and RCAEval~\cite{rcaeval25}), an industrial production deployment.
Averaged over the four backbones, OpsHarness reaches 59.0\% top-1 accuracy, a 63.4\% relative gain over the bare general agent and 4.02× above the specialized agents. 
We further show that OpsHarness continuously improves through self-evolution during the diagnosis process, with each design component contributing to the overall performance gains. Its overhead is comparable to a bare general-purpose agent, while significantly outperforming baseline specialized RCA agents.

This paper makes the following contributions.
\begin{itemize}[leftmargin=*]
\item We propose \textit{OpsHarness}, the first self-evolving external agent harness for RCA. Unlike prior RCA agents built from scratch, OpsHarness focuses on the external layer around diagnosis. It stays compatible with rapidly evolving general agents and reuses their strong general capabilities, while making them better at RCA.
\item We design a self-evolution process that learns from both positive and negative diagnosis trajectories under a dual-gate verification control. It takes the harness from being unfamiliar with a target system to gradually distilling that system's best practices while resisting overfitting, so diagnosis improves the more it is used.
\item We conduct comprehensive experiments on OpsHarness across two public benchmarks and an industrial production deployment. It reaches 59.0\% top-1 accuracy, a 63.4\% relative improvement over a bare general agent and a 4.02$\times$ improvement over baseline RCA agents, and each component contributes to this gain at a reasonable cost.
\end{itemize}

\section{Background and Related Work}
\label{sec:bg}
\textbf{LLM-based RCA.} Classical RCA localizes faults with dependency graphs, causal inference, or spectrum analysis, as in MicroRCA~\cite{microrca20}, Sage~\cite{sage21}, Eadro~\cite{eadro23}, and BARO~\cite{baro24}. With LLMs, recent work builds agents that plan, write code, read telemetry, and reason about the cause. RCAgent~\cite{rcagent24} and the RCA-Agent~\cite{openrca25} act over raw multi-modal telemetry, mABC~\cite{mabc24} and Flow-of-Action~\cite{flowofaction25} coordinate role-specialized agents under SOPs, and others assist incident management from tickets and logs, such as RCACopilot~\cite{rcacopilot24}, Xpert~\cite{xpert24}, and in-context approaches~\cite{ahmed-rcamitigation23, assesssummarize23, iclrca-gpt424, roy-rcaagents24}.
Almost all approaches design a specialized agent from scratch and fit it to a single system, which introduces two key limitations: they cannot leverage rapidly evolving general-purpose agents, and they remain fixed frameworks that do not learn or generalize well to unseen scenarios. OpsHarness, as the first self-evolving external RCA harness, addresses both limitations by being compatible with modern general agents while enabling continual self-evolution during deployment.

\textbf{Harness engineering and general coding agents.} An agent is increasingly understood as a model plus a harness, the external layer that supplies its tools, context, memory, control loop, and evaluation, now studied directly as harness engineering~\cite{agentharness-anatomy26}. CoALA~\cite{coala24} named its components,
and recent surveys formalize the \emph{agent harness} and recast agent building as \emph{externalizing} capability into memory, skills, and tools~\cite{agentharness-survey26, externalization26}. Each component is its own research target, including the agent-computer interface for tools~\cite{sweagent24}, operating-system-style context paging for memory~\cite{memgpt23}, and skill libraries that accumulate reusable procedures~\cite{voyager24, expel24, agentskills25}. In terms of general capability, general coding agents such as Claude Code~\cite{claudecode25}, Codex~\cite{openaicodex25}, and OpenCode~\cite{opencode25} now provide mature tool use, code execution, sandboxing, and long-horizon planning out of the box, and they improve quickly. Their general ability is strong, yet on a specific task they still benefit from a harness layered on top. Prior work typically takes the harness to be everything outside the model; we additionally define \textbf{\emph{external harness}}, the task-specific layer that wraps a whole general agent on top of its general capabilities. For software development, external harnesses such as Superpowers~\cite{superpowers25}, an agentic skills framework, and OMX~\cite{ohmycodex25}, a workflow layer for Codex, already add such an external layer of reusable skills, commands, and workflows in the Dev domain. However, in the Ops domain, such external harnesses are still largely missing. OpsHarness, as the first self-evolving RCA external harness, fills this gap and enables general-purpose agents to achieve stronger capability on RCA tasks through the harness.

\section{Motivation}
\label{sec:mot}
We ground OpsHarness in two observations, drawn from controlled quantitative experiments and from industrial evidence. To obtain solid evidence from a real production setting, we conduct case studies at Company~A, a large commercial cloud service provider, using data from production systems.

\subsection{External Harness is Now the Key to LLM-based RCA}
\label{subsec:harness-key}
General-purpose coding agents already show strong potential on RCA with no RCA-specific design. We run a quantitative experiment across four backbone models (i.e., GPT-5.5, Claude Sonnet~4.6, GLM-5.2, and DeepSeek-V4) and two public benchmarks (i.e., OpenRCA~\cite{openrca25} and RCAEval~\cite{rcaeval25}). For each backbone we compare two settings. In the first, we directly connect the base model to a general agent framework (i.e., Codex or Claude Code). In the second, we connect the same model to a specialized RCA agent designed for the task (i.e., RCA-Agent or mABC). Measured by overall top-1 accuracy across both benchmarks, the general agent is the stronger of the two on every backbone, reaching 36.1\% on average against 17.9\% for RCA-Agent and 5.6\% for mABC (Table~\ref{tab:mot-exp1}). The gap is widest on RCAEval, where the specialized RCA-Agent, designed for OpenRCA, fails to generalize and drop to 5.1\%, while the general agent holds at 38.0\%. 

\begin{table}[t]
\centering
\caption{Top-1 accuracy of General agents vs.\ specialized RCA agents.}
\label{tab:mot-exp1}
\vspace{-0.10in}

\resizebox{\columnwidth}{!}{%
\begin{tabular}{@{}l l ccc@{}}
\toprule
Model & Agent framework & OpenRCA & RCAEval & Overall \\
\midrule
\multirow{3}{*}{GPT-5.5~\cite{openai2026gpt55}}
 & Codex & 44.9 & 57.4 & \textbf{51.2} \\
 & RCA-Agent$^{\dagger}$ & 31.8 & 1.9 & 16.8 \\
 & mABC$^{\dagger}$ & 7.6 & 11.1 & 9.4 \\
\midrule
\multirow{3}{*}{Claude Sonnet 4.6~\cite{anthropic2026sonnet46}}
 & Claude Code & 25.9 & 27.8 & \textbf{26.8} \\
 & RCA-Agent$^{\dagger}$ & 26.4 & 5.6 & 16.0 \\
 & mABC$^{\dagger}$ & 2.2 & 9.3 & 5.8 \\
\midrule
\multirow{3}{*}{GLM-5.2~\cite{glm2026glm5}}
 & Codex & 46.1 & 46.3 & \textbf{46.2} \\
 & RCA-Agent$^{\dagger}$ & 36.8 & 11.1 & 24.0 \\
 & mABC$^{\dagger}$ & 2.2 & 7.4 & 4.8 \\
\midrule
\multirow{3}{*}{DeepSeek-V4~\cite{deepseek2026v4}}
 & Codex & 20.2 & 20.4 & \textbf{20.3} \\
 & RCA-Agent$^{\dagger}$ & 28.0 & 1.9 & 14.9 \\
 & mABC$^{\dagger}$ & 1.2 & 3.7 & 2.5 \\
\bottomrule
\end{tabular}%
}
\vspace{-0.20in}
\end{table}

The reason is that a good RCA agent also depends on the general abilities that flagship agent frameworks have spent years maturing (e.g., context management over long-horizon tasks, multi-round planning, and reasoning). OpenRCA's own analysis of RCA-Agent is telling. Many of its failures stem from gaps in general agentic competence. It is brittle at error recovery and reasons ``lazily'' with short chains~\cite{openrca25}. These are precisely the competencies (i.e., robust tool-use loops, long-horizon planning, and context management) that general agent frameworks now provide out of the box and document as standard practice~\cite{agentharness-anatomy26, sweagent24}. Building an RCA agent from scratch means re-implementing, and usually under-implementing, this machinery.

Yet a bare general agent still falls short of production use. It averages only 36.1\% overall top-1 accuracy across backbones, and even its strongest configuration (GPT-5.5) reaches 51.2\%.
The shortfall is not one of reasoning. A general agent behaves like a broadly capable engineer who is new to the team, in that it brings reasoning ability but lacks diagnostic expertise.

A failure from our general-agent runs makes this concrete. On a Telecom query about a latency spike around 02:15, the agent correctly localizes the anomalous window but incorrectly attributes the root cause to a network-error counter on host \texttt{os\_022}, which shows the largest absolute jump. However, this counter is chronically noisy under load, while the true cause is CPU saturation on container \texttt{docker\_003}, which deviates only modestly in absolute terms from a near-zero baseline. Misled by raw magnitude and lacking awareness of system-specific noise profiles and operational conventions (e.g., near-zero baseline shifts matter more than volatile swings), the agent localizes a downstream symptom rather than the root cause. This reflects missing system-specific diagnostic knowledge rather than a model limitation; such knowledge is exactly what an external RCA-specific harness is designed to provide, aligning general agents with the target system and enabling effective diagnosis.

\begin{findingbox}
\noindent\textbf{Finding 1.} Fast-improving general agents now often outperform previous RCA agents. They already bring strong general diagnostic ability to RCA, but their accuracy is bounded by the external supporting layer.

\smallskip
\noindent\textbf{Implication 1.} The key to LLM-based RCA is to build an external harness around the agent's general capabilities, one that both reuses the flagship agent's mature general capabilities and makes it perform better on RCA.
\end{findingbox}

\subsection{The Ability to Self-Evolve is Decisive for RCA}
\label{subsec:self-evolve}

For a given system, diagnosis is rarely a one-shot exercise but an ongoing process, as similar incidents often recur over a system’s operational lifetime (e.g., 36.2\% of incidents with recorded troubleshooting guides occur at least twice~\cite{tsgrec20}). What distinguishes expert SREs from capable newcomers is the ability to continuously summarize best practices and mine experience from each diagnosis, i.e., to self-evolve. This self-evolving capability enables an agent to transition from unfamiliarity with a newly onboarded system to progressively extracting expertise and improving its performance over time, achieving true generalization across systems.

\begin{figure}[t]
\centering
\includegraphics[width=\linewidth]{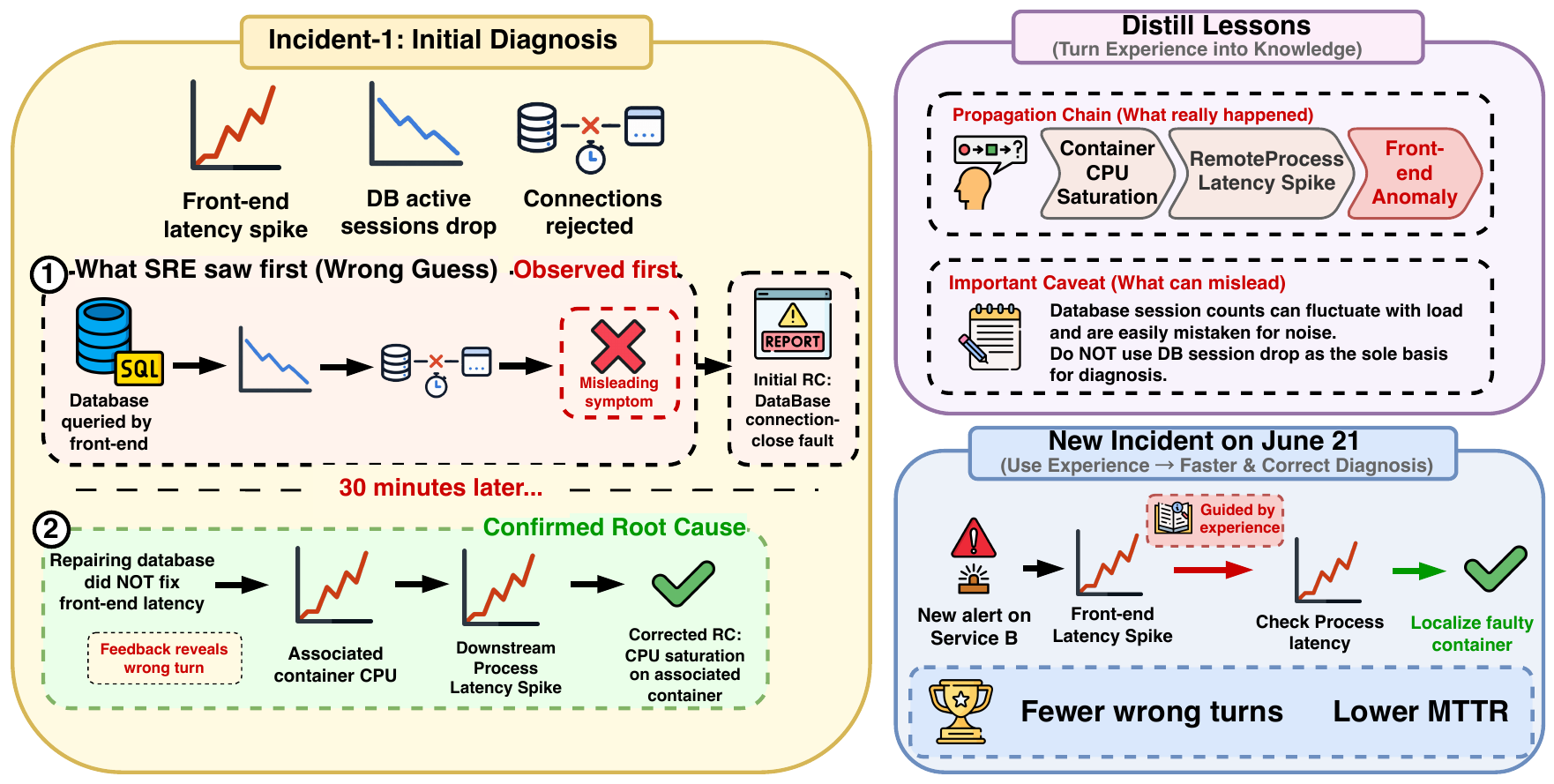}
\vspace{-0.30in}
\caption{A continuous diagnosis process on one system, in which an expert SRE accumulates and reuses best practices across incidents.}
\vspace{-0.20in}
\label{fig:casestudy}
\end{figure}

We illustrate this with a real-world production case from Company~A (Figure~\ref{fig:casestudy}). \textbf{(1)}~On June~15, an on-call SRE received a change-induced alert for service $\alpha$, where front-end latency spiked and front-end containers showed resource saturation. The SRE observed that database active sessions dropped sharply from 391 to 52, with connections rejected within about 40~ms, and attributed the root cause to a database connection-close fault. However, 30 minutes later, downstream engineer reported that fixing the database did not resolve the issue. Further investigation showed a different associated container where CPU utilization surged from 1.6\% to 98\%, triggering a downstream RemoteProcess latency spike; the true root cause was CPU saturation on that container. After restart, all services recovered, and the database session drop was identified as incidental noise. \textbf{(2)}~From this diagnosis, the SRE recorded two lessons: the propagation chain from container CPU saturation to RemoteProcess latency spike and front-end anomaly, and the caveat that database session counts can fluctuate under load and are not reliable standalone signals. \textbf{(3)}~On June~21, the same SRE handled another alert on service $\beta$. Guided by prior experience, the SRE directly checked RemoteProcess latency and container CPU, quickly localized the faulty container, and resolved the incident via restart. This case demonstrates that continuously distilling and reusing diagnostic experience is critical in production, and that expert SRE accuracy stems from such accumulated knowledge, which a strong RCA agent should also be able to self-evolve.

Most operational knowledge and its evolution are still created manually. Engineering teams document expert experience and recurring fixes as Troubleshooting Guides (TSGs). At Microsoft, over 4,000 TSGs are linked to thousands of incidents, and attaching the right TSG reduces mean time to mitigate a severity-2 incident from about 18 to 13 hours, yet TSGs are still written and executed manually~\cite{autotsg22}. A study of 152 high-severity incidents shows that over 90\% are mitigated using such procedures~\cite{fightincidents22}, while deeper root-cause knowledge remains in unstructured postmortems that are “not directly reusable”~\cite{saha-rckg22, assesssummarize23}. Overall, the artifacts defining SRE expertise—TSGs, diagnostic skills, and fault-propagation patterns—are learned from past incidents but largely maintained manually or in practitioners’ heads.

This motivates an evolving harness that automatically mines each diagnosis for reusable best practices, knowledge, and tools, and applies them to future diagnoses on the same system.

\begin{findingbox}
\noindent\textbf{Finding 2.} Both SREs and agents need self-evolving to mine expert experience from diagnostic cases and reuse it in subsequent diagnoses. However, this evolution process and the resulting expertise are still largely manual today.

\smallskip
\noindent\textbf{Implication 2.} 
Self-evolution is a core capability of an RCA harness, which enables an agent to generalize to unfamiliar systems, progressively extract system-specific expertise during diagnosis, and continuously improve its performance to achieve effective analysis.
\end{findingbox}

\section{OpsHarness Design}
\label{sec:design}

\subsection{Overview}
\label{subsec:design-overview}
\begin{figure*}[t]
\centering
\includegraphics[width=1\linewidth]{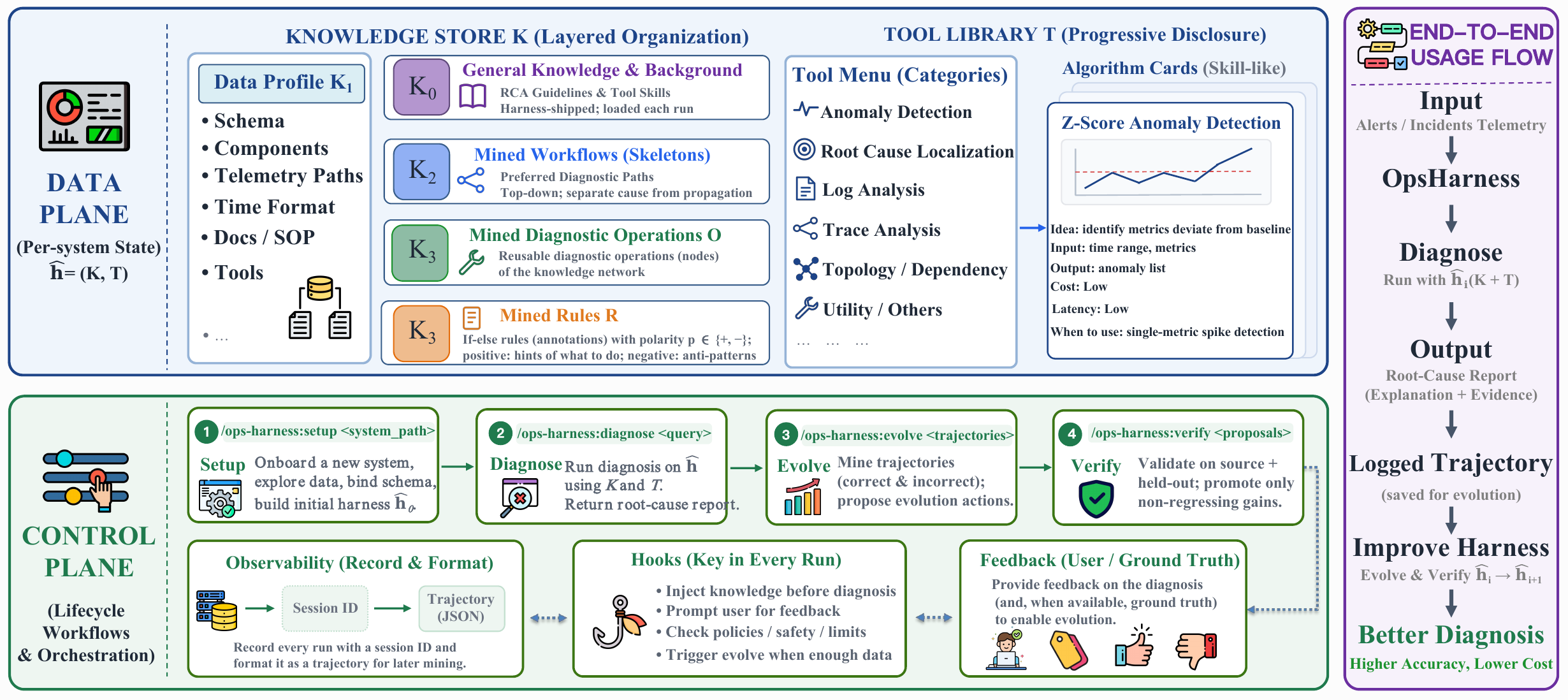}
\vspace{-0.20in}
\caption{An overview of OpsHarness design.}
\vspace{-0.20in}
\label{fig:overview}
\end{figure*}

We organize OpsHarness into a \textbf{data plane} and a \textbf{control plane}, as shown in Fig.~\ref{fig:overview}. The data plane is the per-system state that OpsHarness adapts and evolves. We write it as
\begin{equation}
\label{eq:state}
h = (\mathcal{K},\, \mathcal{T}),
\end{equation}
a layered operational knowledge store $\mathcal{K}$ (Section~\ref{subsec:knowledge}) and an idea-card tool library $\mathcal{T}$ (Section~\ref{subsec:tools}). The control plane is a set of lifecycle workflows over $h$, namely \textit{setup}, \textit{diagnose}, \textit{evolve}, and \textit{verify}, together with an observability subsystem that records each run. Setup and diagnose adapt and apply the harness (Section~\ref{subsec:lifecycle}), while evolve and verify form the self-evolution loop that advances it from one version to the next, $h_i \to h_{i+1}$ (Section~\ref{sec:evolve}).


\subsection{Layered Operational Knowledge}
\label{subsec:knowledge}
Operational knowledge is not flat. Some of it is experience about how a diagnosis should flow, while some is best practice for one concrete operation step. OpsHarness therefore organizes knowledge into four tiers,
\begin{equation}
\label{eq:knowledge}
\mathcal{K} = (\mathcal{K}_0,\, \mathcal{K}_1,\, \mathcal{K}_2,\, \mathcal{K}_3),
\end{equation}
each with its own disclosure and update policy.

$\mathcal{K}_0$ stores \textbf{general knowledge and background} about the RCA process. At a high level, it defines what a diagnosis should do and what it should produce. $\mathcal{K}_0$ ships with the harness as text in its root documents (e.g., \texttt{AGENTS.md} or \texttt{CLAUDE.md}) and is loaded on every diagnosis.

$\mathcal{K}_1$ stores \textbf{a profile of the target system} that captures its metadata (e.g., its schema, on-disk layout, and component inventory). $\mathcal{K}_1$ is produced automatically once \textit{setup} completes, and it is read in full at the start of each diagnosis. It replaces a per-dataset adapter, in that it tells the agent how to read this system without any hard-coded loader.

$\mathcal{K}_2$ and $\mathcal{K}_3$ are both mined from diagnosis trajectories during evolution (Section~\ref{sec:evolve}). Together they form the target system's diagnostic knowledge network, the automated counterpart of the expertise an SRE accumulates over a system. We organize this network as a directed graph $G = (\mathcal{O}, E, \mathcal{R})$. The operations $\mathcal{O}$ are the nodes. A directed edge $(o, o') \in E$ records that $o'$ is a sensible next move after $o$ on this system. Each rule in $\mathcal{R}$ is a typed annotation on a diagnostic state. A workflow skeleton is a path in $G$. The two mined tiers slice this graph by granularity. The coarse workflow paths form $\mathcal{K}_2$, and the fine-grained nodes and annotations form $\mathcal{K}_3$,
\begin{equation}
\label{eq:k2}
\mathcal{K}_2 = \Pi \subseteq \mathrm{Paths}(G), \qquad \mathcal{K}_3 = (\mathcal{O},\, \mathcal{R}).
\end{equation}


\noindent\textbf{$\mathcal{K}_2$, mined workflows.} A skeleton $\pi = (o_1, \dots, o_m)$ is an ordered workflow over operations, i.e., a path through the network that represents one of the system's preferred diagnosis workflows (e.g., confirm the front-end KPI window, localize top-down across layers, then separate cause from propagation).

\noindent\textbf{$\mathcal{K}_3$, mined operations and rules.} The fine-grained tier, $\mathcal{K}_3 = (\mathcal{O}, \mathcal{R})$. Operations $\mathcal{O}$ are reusable diagnostic moves, the nodes of the network (e.g., isolate container from node CPU, or pivot on the caller of a failed trace). Rules $\mathcal{R}$ are fine-grained if-else diagnostic rules, the annotations of the network. Each rule
\begin{equation}
\label{eq:rule}
r = (\phi,\, \psi,\, p), \qquad p \in \{+, -\},
\end{equation}
pairs a condition $\phi$ with an action or caveat $\psi$ and a polarity $p$. A positive rule is a hint of what to do (e.g., a container CPU surge propagates to a RemoteProcess latency spike, then to the front end). A negative rule is an anti-pattern, a thing not to do (e.g., database session counts fluctuate with load and should not be the sole basis for a diagnosis).

\noindent\textbf{Progressive disclosure.} OpsHarness stores all knowledge as text under the harness root. Loading every tier on each diagnosis would crowd out the agent's context and degrade its reasoning, so disclosure is progressive. The small, always-relevant $\mathcal{K}_0$ and $\mathcal{K}_1$ are loaded on every diagnosis. Each $\mathcal{K}_2$ workflow carries a short summary; a diagnosis first loads these summaries, then loads in full only the workflow it selects. That workflow in turn drives recall in $\mathcal{K}_3$, pulling in the relevant operations and rules on demand as the diagnosis reaches a state to which they apply. 

\subsection{The Idea-Card Tool Library}
\label{subsec:tools}
A diagnosis usually needs analytical tools (e.g., anomaly detection, feature scoring, and suspect ranking). OpsHarness organizes them as idea cards rather than pre-written scripts. An \textit{idea card} is a skill that gives a natural-language specification of one algorithm, stating its core idea, its procedure as light pseudocode, its input and output contract, when to use it, and its key parameters. At diagnosis time the agent reads the card and writes the few lines of code that implement it against the active schema. We make this choice because operational data layouts differ across systems and even across windows, so a once-generated hard-coded script tends to break on a schema it did not anticipate, whereas code the agent writes and iterates against the live context is more likely correct for the diagnosis at hand.
An operation in $\mathcal{K}_3$ (Section~\ref{subsec:knowledge}) often maps to a tool idea card here, pairing a diagnostic move with the algorithm that realizes it.

The library is disclosed in the same staged way: the agent first reads a one-page menu of algorithms with a short ``use when'', then opens only the cards it needs, never the whole library. At cold start it holds only scaffolding placeholders and no system-specific tools; as diagnoses accumulate, evolve distills the fixed code logic that recurs across correct trajectories into new tool cards, so the library gradually fills with diagnostic tools tuned to the target system.

\subsection{The Diagnosis Lifecycle}
\label{subsec:lifecycle}
We detail \textit{setup} and \textit{diagnose} here, since \textit{evolve} and \textit{verify} form the self-evolving loop of Section~\ref{sec:evolve}. 

\noindent\textbf{Setup.} Setup is a skill that profiles the target system, exposed as a slash command (i.e., \texttt{/opsharness:setup}) and run when OpsHarness first onboards a system or when its structure changes. The user only points OpsHarness at the telemetry (e.g., a path on disk, or a database endpoint). OpsHarness then queries and samples the data, detects the semantic columns from repeated samples (e.g., entity, metric name, and metric value), builds the entity inventory and metric categories, and writes $\mathcal{K}_1$ together with a profile YAML, all with no per-dataset code. If the system already ships with knowledge or tools, the user can pass their access paths, and setup folds what it finds into $\mathcal{K}_1$.


\begin{figure*}[t]
\centering
\includegraphics[width=0.9\linewidth]{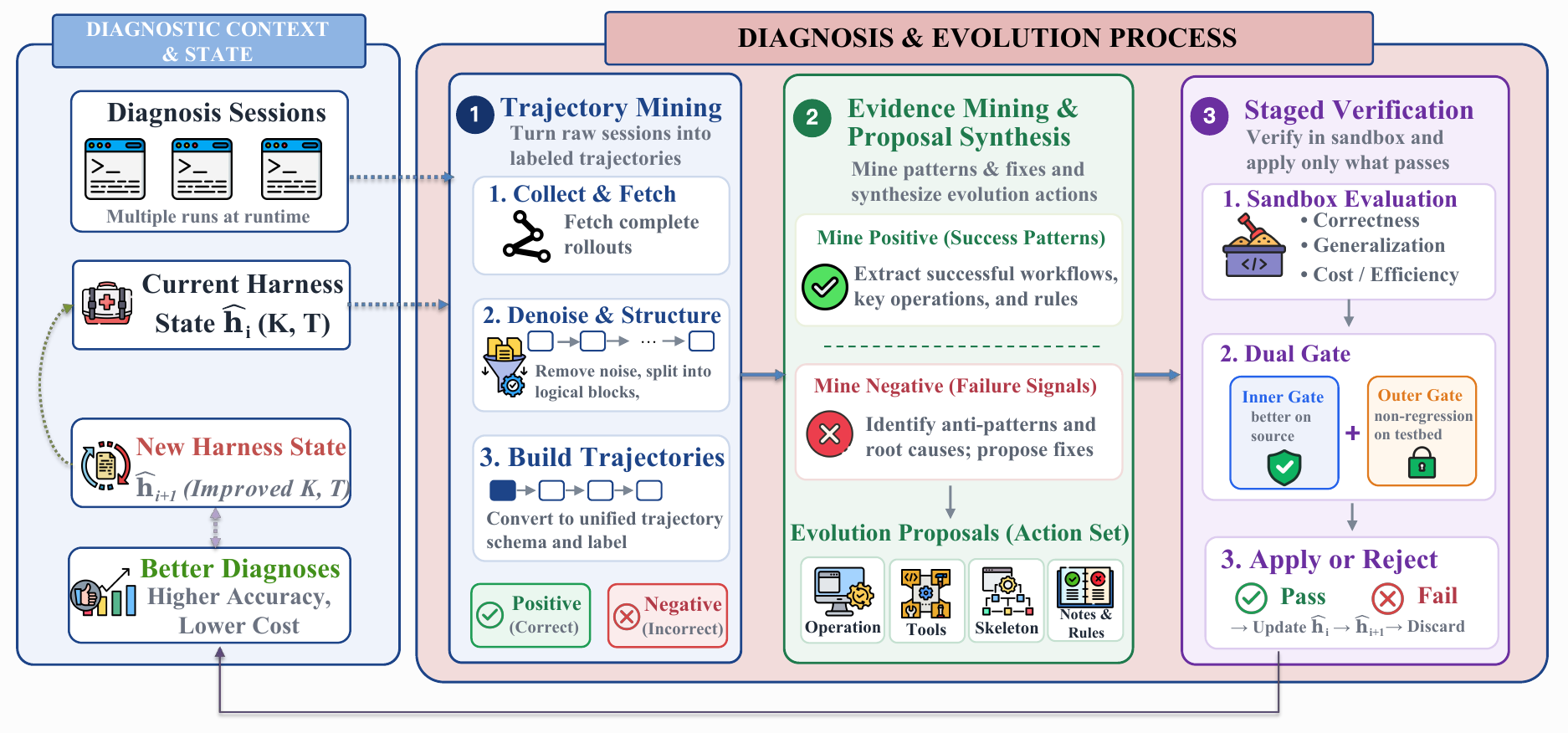}
\vspace{-0.10in}
\caption{An overview of the self-evolving loop in OpsHarness.}
\vspace{-0.20in}
\label{fig:evolve}
\end{figure*}

\noindent\textbf{Diagnose.} Diagnose is likewise a slash command (i.e., \texttt{/opsharness:diagnose \textless query\textgreater}) that the user invokes or that an alert triggers. Given a diagnosis query or a symptom alert, a diagnosis proceeds top-down. The agent loads $\mathcal{K}_0$ and $\mathcal{K}_1$, consults $\mathcal{K}_2$ and $\mathcal{K}_3$ where relevant, and narrows \textit{when} (the anomaly window), \textit{where} (the responsible layer and component), and \textit{why} (the fault type), cross-validating across modalities (e.g., metrics confirm, logs explain, and traces show the impact path). Along the way it implements the idea cards it needs in code against the current diagnostic context and runs them. The output is a ranked root-cause report (i.e., a component, a fault type, and an onset time), and the observability subsystem captures the run as a trajectory. A diagnosis thus uses the agent's general capabilities as connective tissue, assembling the diagnosis-specific knowledge and tools that setup and evolve prepared to carry it through.

\section{Self-Evolving in OpsHarness}
\label{sec:evolve}


The self-evolving process runs in three stages (Fig.~\ref{fig:evolve}). \textit{Trajectory mining} turns raw sessions into labeled diagnosis trajectories (Section~\ref{subsec:mining}). \textit{Evidence mining and proposal synthesis} summarizes the successful patterns in the positive trajectories and the actionable fixes in the negative ones, emitting a set of evolution actions (Section~\ref{subsec:proposal}). \textit{Staged verification} runs each proposal through a dual gate in a sandbox and applies only the ones that pass (Section~\ref{subsec:verify}).

\subsection{Trajectory Mining}
\label{subsec:mining}

\noindent\textbf{Diagnosis trajectory.} A session is a time-ordered sequence of events, and slash-command markers delimit its activity. A \textit{diagnosis block} is a maximal span that opens at a \textit{diagnose} command and closes at its end marker. Each tracked block becomes one trajectory $\tau$, while the spans outside OpsHarness logic are discarded as noise.

After a diagnosis is resolved, a feedback hook fires and asks the user for lightweight feedback. The user gives a thumbs-up or thumbs-down and the actual root cause for that diagnosis from downstream. Trajectory mining then reads the feedback for each diagnosis and attaches a label to its trajectory,
\begin{equation}
\label{eq:label}
\ell(\tau) = (a,\, c),
\end{equation}
the accuracy $a \in \{\text{correct}, \text{partial}, \text{wrong}\}$ derived from the feedback, and the cost $c$ (i.e., tokens and time) measured by observability. The labels split the batch into a positive and a negative set,
\begin{equation}
\label{eq:split}
\mathcal{D}^{+} = \{\tau : a(\tau) = \text{correct}\}, \ \mathcal{D}^{-} = \{\tau : a(\tau) = \text{wrong}\}.
\end{equation}

\subsection{Evidence Mining and Proposal Synthesis}
\label{subsec:proposal}
The evolver mines $\mathcal{D}^{+}$ and $\mathcal{D}^{-}$ into a proposal. We define a proposal as a set of atomic, typed operators over the data-plane state. Each operator is one element of
\begin{equation}
\label{eq:operators}
\omega \in \{\mathrm{add}, \mathrm{update}, \mathrm{delete}\} \times \{\textsc{skel}, \textsc{op}, \textsc{rule}, \textsc{tool}\},
\end{equation}
a proposal is a set $P = \{\omega_1, \dots, \omega_k\}$, and applying it gives a candidate $h' = \mathrm{Apply}(h_i, P)$. An operator edits exactly one element, namely a skeleton path $\pi$ in $\mathcal{K}_2$, an operation or a rule in $\mathcal{K}_3$, or a tool card in $\mathcal{T}$, and thus advances the full data-plane state $h=(\mathcal{K},\mathcal{T})$. 

\noindent\textbf{Learning from correct runs (commonality).} Correct trajectories share structure worth making explicit and reusable. The evolver extracts this commonality at several granularities. A recurring whole-diagnosis order becomes or refines the skeleton $\pi$. A move that recurs across correct runs becomes an operation node (e.g., isolating container from node CPU appeared in every correct CPU diagnosis), and a block of fixed code that recurs across them is distilled into a tool card in $\mathcal{T}$ (Section~\ref{subsec:tools}). A recurring observation becomes a positive rule. Conceptually, an element $o$ is promoted when it recurs across a sufficient fraction of correct trajectories,
\begin{equation}
\label{eq:commonality}
\mathrm{add}\ o \quad \text{when} \quad \mathrm{freq}(o,\, \mathcal{D}^{+}) \ge \rho,
\end{equation}
a criterion the evolver approximates by reasoning over the labeled batch. The recurring order is distilled into a $\mathcal{K}_2$ skeleton, pruned to its decisive steps so that later diagnoses replay the efficient path rather than rediscover it, dropping the exploratory detours that made early runs expensive.

\noindent\textbf{Learning from incorrect runs (contrastive correction).} A wrong trajectory reveals where reasoning diverges from the truth. The evolver identifies the \textit{divergence point}—the first step inconsistent with the confirmed root cause—by comparing every step's intent of the trajectory against the ground truth, and converts the resulting corrective difference into an edit. When a correct and an incorrect diagnosis of similar incidents are available, the skipped moves after their divergence point become the fix. Misleaders recurring across wrong runs are distilled into negative rules, while failures caused by outdated evolved knowledge trigger updates or deletions.

\subsection{Staged Dual-Gate Verification}
\label{subsec:verify}
A mined proposal is a hypothesis, not yet an improvement. It may overfit its source cases, or help locally while hurting elsewhere. OpsHarness admits a change only after it passes two gates in an isolated sandbox.

\noindent\textbf{Sandbox.} The verifier creates a \textit{stage}, a pristine copy of the live harness, and applies $P$ there to obtain $h' = \mathrm{Apply}(h_i, P)$. The live harness is untouched until promotion.

\noindent\textbf{Inner gate: strictly better on the source cases.} The evolved harness must first beat the original harness on the cases it was mined from. For a harness $h$ and a case set $D$, let $A_h(D)$ be the accuracy rate and $C_h(D)$ the mean cost (in tokens). On the source batch $\mathcal{D}_i$ we measure $h_i$ and $h'$ and compare them. With deltas $\Delta A = A_{h'}(\mathcal{D}_i) - A_{h_i}(\mathcal{D}_i)$ and $\Delta C$ on $\mathcal{D}_i$, and a cost tolerance $\beta = 0.05$,
\begin{equation}
\label{eq:okpreds}
\mathrm{accOK} \equiv \Delta A \ge 0, \mathrm{costOK} \equiv \Delta C \le \max(\beta\, C_{h_i},\, 1),
\end{equation}
\begin{equation}
\label{eq:gin}
g_{\mathrm{in}} \equiv \mathrm{accOK} \wedge \mathrm{costOK} \wedge (\Delta A > 0 \vee \Delta C < 0).
\end{equation}
The inner gate requires no regression in accuracy, no material increase in cost, and a strict improvement in at least one objective. The strict clause keeps out no-op changes, and the cost band tolerates a small token increase but rejects accuracy bought with runaway exploration.

\noindent\textbf{Outer gate: non-regression on a held-out testbed.} Passing on the source cases does not show that a change generalizes. The verifier maintains a separate testbed $\mathcal{B}$, kept diverse over time and over fault family (e.g., CPU, memory, disk, network, and database). Rather than being fixed upfront, $\mathcal{B}$ is maintained at run time: before every evolution, we cluster the accumulated cases in parallel by fault family and by occurrence time and draw a stratified sample, sized by default to the evolve window. Resampling ahead of each evolve step keeps $\mathcal{B}$ current and disjoint from the mined cases, so a proposal cannot be fit to a static held-out set across iterations. Each testbed case is stored as a truth-free task prompt with a privately held label, and the baselines are refreshed whenever the live harness changes, so the bar is always current. On $\mathcal{B}$ we require only non-regression,
\begin{equation}
\label{eq:gout}
g_{\mathrm{out}} \equiv \mathrm{accOK} \wedge \mathrm{costOK} \quad (\text{measured on } \mathcal{B}).
\end{equation}

\noindent\textbf{Acceptance and refinement.} A change is promoted if and only if both gates pass,
\begin{equation}
\label{eq:accept}
\mathrm{accept} \equiv g_{\mathrm{in}}(h', h_i;\, \mathcal{D}_i) \wedge g_{\mathrm{out}}(h', h_i;\, \mathcal{B}),
\end{equation}
in which case $h_{i+1} \leftarrow h'$ atomically. If a gate fails, the proposal returns to the evolver, which refines it with the failing cases as fresh negatives and retries, up to three rounds. A proposal that still does not pass is rejected and discarded, and its stage environment is torn down to reclaim space.


\section{Evaluation}
\label{sec:eval}
\begin{itemize}[leftmargin=*]
\item \textbf{RQ1:} How effective is OpsHarness, especially after self-evolution?
\item \textbf{RQ2:} How much does each design of \textit{OpsHarness} contribute?
\item \textbf{RQ3:} What is the cost of OpsHarness?
\item \textbf{RQ4:} How does OpsHarness perform on an industrial deployment?
\end{itemize}

\subsection{Experimental Setup}
\label{subsec:setup}

\subsubsection{Datasets}
We evaluate on two public RCA benchmarks and one industrial dataset. The two public benchmarks together cover a broad range of systems, fault types, and noise profiles.

\noindent\textbf{OpenRCA}~\cite{openrca25} contains 335 real-world cases drawn from 3  enterprise software systems, namely a telecom system (Telecom), a banking system (Bank), and a market system (Market). 
The dataset involves 46 nodes, 68 containers, and 176 service meshes, while covering 28 distinct root-cause categories, providing comprehensive evaluation across systems of different scales and diverse fault types.

\noindent\textbf{RCAEval}~\cite{rcaeval25} contains 270 labeled cases over 3 widely used open-source microservice benchmarks, namely Online Boutique, Sock Shop, and Train Ticket. Faults are injected with chaos engineering and cover CPU, memory, network, disk, and code-level anomalies. 
RCAEval involves 97 monitored entities and 87 microservices, also spanning all 3 telemetry modalities.

\input{tabs/exp_1}

\noindent\textbf{Industrial dataset.} We additionally use a production change-anomaly dataset from Company~A. The telemetry is reported through Prometheus and persisted to a distributed log store. The dataset covers 773,340 data points and 88 confirmed anomalies that occurred during change windows. Labels come from on-call SREs feedback combined with the pulled-back raw logs.


\noindent\textbf{Train/test split.} For every (sub)dataset we hold out the last 20\% of cases in time order as the test set and use the earlier 80\% for warm-up (i.e., self-evolution for OpsHarness and demonstration retrieval for in-context-learning baselines). A temporal split mimics a real deployment, in which a system is diagnosed going forward in time, and it prevents any method from seeing future cases. All numbers below are on the held-out test set unless stated otherwise.

\subsubsection{Baselines and Implementation}
We evaluate every method as a (backbone model $\times$ agent framework) pair.

\noindent\textbf{Backbone models.} We use four recent models spanning open and closed weights and a range of cost and capability, namely GPT-5.5~\cite{openai2026gpt55}, Claude Sonnet~4.6~\cite{anthropic2026sonnet46}, GLM-5.2~\cite{glm2026glm5}, and DeepSeek-V4~\cite{deepseek2026v4}, the latest releases at the time of experiments. We exclude very expensive models (e.g., Claude Opus), as they are hard to run at  volume a production RCA pipeline requires, whereas our goal is a setting operators can deploy.

\noindent\textbf{Agent frameworks.} We compare six frameworks, two specialized and four built on a general agent. \textbf{RCA-Agent}~\cite{openrca25}, the specialized agent proposed with OpenRCA, plans, writes and runs code over telemetry, and reasons about the cause, while \textbf{mABC}~\cite{mabc24} is a multi-agent framework whose role-specialized agents collaborate to localize the cause. \textbf{Direct} connects the backbone to a general agent framework (i.e., Claude Code or Codex) with no RCA harness, and \textbf{ICL} adds few-shot in-context learning~\cite{icl} to it, retrieving three correct rollouts of the most similar past queries into the prompt. \textbf{OpsHarness (no-evolve)} is OpsHarness with self-evolution disabled (i.e., the Day-0 substrate only), measuring the harness's cold-start ability, and \textbf{OpsHarness} is our full method with self-evolution enabled.
We pair Claude Sonnet~4.6 with Claude Code and the other three models with Codex, following each vendor's supported agent. 
For OpsHarness, we integrate it into the corresponding general-purpose agent.
We run every method with the same environment and available information. 

\subsubsection{Evaluation Metrics}
We report effectiveness and efficiency metrics standard in RCA evaluation~\cite{baro24, openrca25}. A diagnosis is scored against the labeled root cause, i.e., a component and, where applicable, a fault type and onset time. \textbf{A@$k$} (top-$k$ exact accuracy, $k\in\{1,3\}$) is the fraction of cases whose complete root-cause tuple appears among the top-$k$ candidates. \textbf{Avg Score} is the mean per-case score giving partial credit across root-cause elements (e.g., the component is correct but the fault type differs). For efficiency we report the average tokens and time per case.

\subsection{RQ1: Effectiveness}
\label{subsec:rq1}

To answer RQ1, we run all 24 instances on the two public benchmarks and report A@1, A@3, and Avg Score on each of the six sub-datasets, plus a Final A@1, defined as the mean of the A@1 values (Table~\ref{tab:rq1-main}). Three results stand out.

\noindent\textbf{\textit{OpsHarness is the strongest framework on every backbone.} }Averaged over the four backbones, full OpsHarness reaches 59.0\% Final A@1, against 41.4\% for OpsHarness (no-evolve), 38.4\% for ICL, and 36.1\% for Direct, while the specialized RCA-Agent and mABC reach only 17.9\% and 5.6\%. The two gains compound. The cold-start substrate lifts a bare agent by 5.3 points at cold start (Direct 36.1\% to no-evolve 41.4\%), and self-evolution adds a further 17.6 points (to 59.0\%), for a total of 22.9 points over the bare agent. 

\noindent\textbf{\textit{The cold-start substrate helps most on weaker backbones, and ICL is not enough.}} On average the cold-start substrate beats both Direct (41.4\% vs 36.1\%) and ICL (38.4\%), but the picture varies by backbone. On the three weaker backbones (GLM-5.2, DeepSeek-V4, and Claude Sonnet~4.6) it clearly beats Direct, with the largest lift on the weakest model (DeepSeek-V4, 20.3\% to 31.6\%). On the strongest backbone (GPT-5.5) the bare agent is already strong (Direct 51.2\%), so the generic substrate adds little at cold start (no-evolve 52.8\%) and the harness's edge there comes almost entirely from self-evolution. ICL helps only marginally on average (38.4\% vs 36.1\% for Direct) and does not improve overall Avg Score (67.0\% vs 68.8\% for Direct), because retrieving similar rollouts injects raw examples but not distilled, reusable best practices. This matches Finding~2, that what compounds is mined experience, not raw recall of past cases.

\noindent\textbf{\textit{Specialized agents trail the general agents.}} On the same backbone, the from-scratch RCA-Agent and mABC sit well below even Direct (17.9\% and 5.6\% vs 36.1\% Final A@1), echoing the motivation study (Section~\ref{subsec:harness-key}). 

\subsection{RQ2: Contribution of Each Design}
\label{subsec:rq2}

\begin{figure}[t]
\centering
\includegraphics[width=\linewidth]{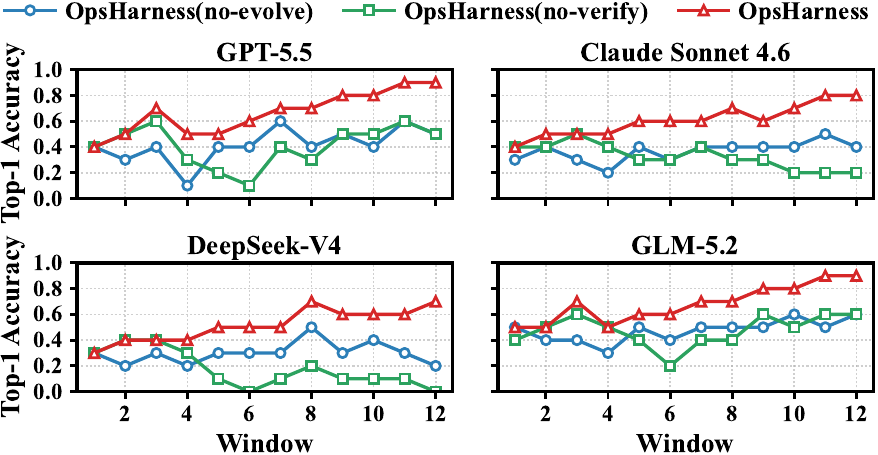}
\vspace{-0.25in}
\caption{Per-window A@1 over 12 continuous-diagnosis windows for three OpsHarness variants on four backbones. }
\vspace{-0.20in}
\label{fig:rq2-curve}
\end{figure}

To isolate the contribution of self-evolution and of its verification gate, we run three variants of OpsHarness in continuous-diagnosis mode over the training split of both benchmarks: the full system, \textbf{OpsHarness (no-evolve)} (the cold-start substrate with self-evolution disabled), and \textbf{OpsHarness (no-verify)}(self-evolution on but the verification gate removed, promoting any proposal that helps its source cases). Each variant diagnoses the cases in order; we split the stream into 12 windows, trigger one evolution per window, and report the mean A@1 per window. We repeat this for all four backbones (Fig.~\ref{fig:rq2-curve}).

\subsubsection{Self-Evolution Improves with Use}
On every backbone, full OpsHarness climbs steadily across the 12 windows, roughly doubling its A@1 from the first window to the last 
(e.g., 0.4 to 0.9 on GPT-5.5, 0.5 to 0.9 on GLM-5.2), 
and ends as the most accurate variant in all four panels. OpsHarness (no-evolve) stays essentially flat (final-window A@1 around 0.2 to 0.6), since a static harness cannot learn the target system. Averaged over the four backbones, the final-window A@1 is 0.83 for full OpsHarness against 0.43 for no-evolve.
This gain reflects the system expertise the harness accumulates with use.

\subsubsection{The Verification Gate Prevents Overfitting}
The no-verify variant shows why the gate matters. With self-evolution on but unverified, accuracy first rises as early proposals help, but then stalls or regresses: averaged over the four backbones its final-window A@1 is 0.33, below even the non-evolving harness (0.43). The damage concentrates on the weaker backbones, whose proposals overfit most. On DeepSeek-V4 no-verify collapses from a peak of 0.4 to 0.0.
A weaker model is more likely to mine a spurious pattern from its source cases. For example, in one cycle the evolver saw repeated failures on network-type cases and proposed up-weighting network anomalies during reasoning; this helped the source cases but, in the next window, mislabeled CPU- and memory-type cases that were previously correct, because the no-verify evolution now surfaced network noise as the cause. The outer gate rejects exactly such proposals: it promotes a change only if it does not regress on a disjoint held-out testbed. With the gate on, every promoted change holds or improves out of sample, which is why full OpsHarness keeps climbing while no-verify decays. 
On average, 37\% of evolution proposals are rejected by the verification gate, and accepted proposals require 1.7 attempts.

\subsubsection{Knowledge and Tool Mechanisms}
To report how the harness learns, we characterize the quality of  artifacts it evolves and how often they are used (Table~\ref{tab:rq2-knowledge}). 
For each family we report the count, the fraction of test cases that recall it (Recall), and the fraction of correct cases in which it was recalled (Precision). We can see higher-granularity knowledge is recalled broadly, while fine-grained note-rules and tools are recalled selectively but with high effectiveness when they fire.

\begin{table}[t]
\centering
\caption{RQ2: evolved knowledge and tools on one system.}
\vspace{-0.10in}
\label{tab:rq2-knowledge}
\resizebox{0.8\columnwidth}{!}{%
\begin{tabular}{@{}l c c c@{}}
\toprule
Artifact family & Count & Recall & Precision  \\
\midrule
Workflow skeleton   &  6 & 91.7 & 73.5 \\
Operation skill     & 19 & 82.5 & 84.3 \\
Note-rule           & 21 & 64.7 & 85.1 \\
Persisted tool      & 11 & 61.0 & 77.2 \\
\bottomrule
\end{tabular}%
}
\vspace{-0.10in}

\end{table}

\subsection{RQ3: Overhead}
\label{subsec:rq3}

\begin{figure}[t]
\centering
\includegraphics[width=\linewidth]{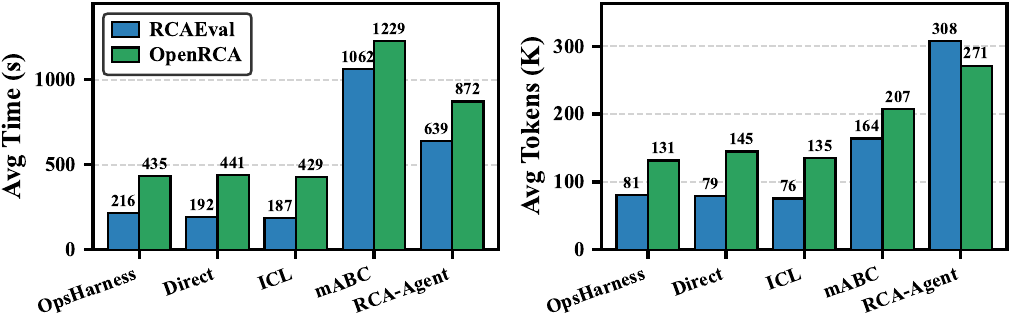}
\vspace{-0.25in}
\caption{Per-case diagnosis cost on the two benchmarks.}
\label{fig:rq3-cost}
\end{figure}

\subsubsection{Diagnosis Cost}
We compare the per-case Avg Token and Avg Time of OpsHarness against Direct, ICL, and the two specialized agents (Fig.~\ref{fig:rq3-cost}). Although OpsHarness loads skills and knowledge into context, it cuts trial-and-error during diagnosis, so its cost stays in line with the lightweight baselines. Averaged over both benchmarks, it spends about 106k tokens and 325s per case, on par with Direct (112k, 317s) and ICL (106k, 308s). The specialized agents cost far more, about 1.7$\times$ to 2.7$\times$ the tokens and 2.3$\times$ to 3.5$\times$ the wall-clock of OpsHarness 
, mostly because they retry and re-plan more.

\subsubsection{Setup, Evolution, and Verification Cost}
Beyond per-case diagnosis, OpsHarness pays cost in three non-diagnosis stages: setup (once per dataset), evolution, and verification (once per cycle). On average, one pass costs 0.82M tokens and 412s for setup, 0.75M and 184s for evolution, and 1.55M and 421s for verification (Fig.~\ref{fig:rq3-footprint}). Verification dominates the token bill (49.8\%) because it runs held-out diagnoses in parallel, while evolution is the lightest stage. These stages run rarely, executed in parallel as sidecar, and amortize over many later diagnoses, so the recurring cost is the per-case diagnosis cost above.

\subsubsection{Harness Footprint}
\begin{figure}[t]
\centering
\includegraphics[width=\linewidth]{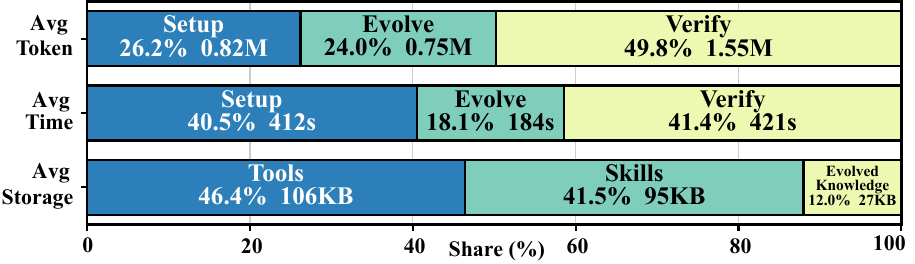}
\vspace{-0.25in}
\caption{OpsHarness per-stage cost, and on-disk footprint by component.}
\vspace{-0.20in}
\label{fig:rq3-footprint}
\end{figure}

Finally, we report the harness's on-disk footprint after warm-up (Fig.~\ref{fig:rq3-footprint}, Avg Storage). It totals about 228KB: reusable tools (106KB, 46.4\%), the skill library (95KB, 41.5\%), and the evolved knowledge learned for the target system (27KB, 12.0\%). 
The whole footprint is negligible relative to the per-case telemetry.

\subsection{RQ4: Industrial Deployment}
\label{subsec:rq4}

\subsubsection{Results on the Industrial Dataset}
\begin{figure}[t]
\centering
\includegraphics[width=\linewidth]{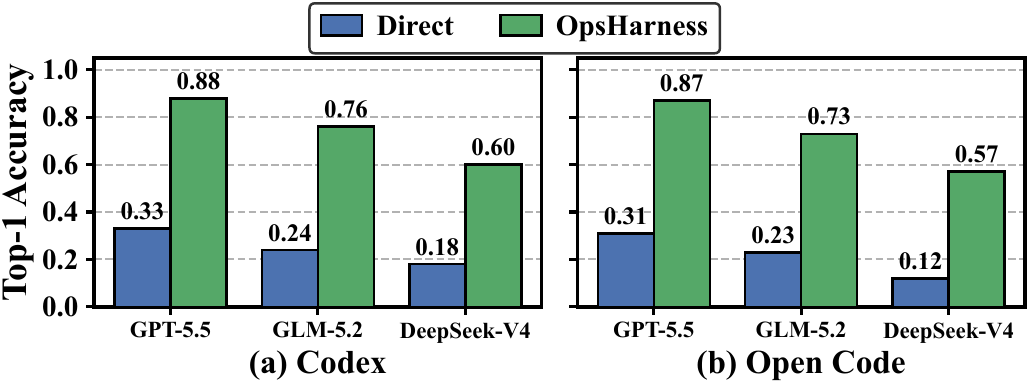}
\vspace{-0.25in}
\caption{Industrial effectiveness on the Company~A dataset.}
\vspace{-0.25in}
\label{fig:rq4-industrial}
\end{figure}

We deploy OpsHarness on the Company~A change-anomaly dataset and compare it to Direct under Codex and open-source Open Code with 3 models, reporting A@1 (Fig.~\ref{fig:rq4-industrial}). OpsHarness improves over Direct in all six configurations. Averaged over them it reaches 0.74 A@1 against 0.24 for Direct, roughly a 3$\times$ lift. The gains do not depend on a proprietary stack: on the open-source Open Code agent, OpsHarness still reaches 0.73 A@1 with the open-source GLM-5.2 and 0.57 with DeepSeek-V4 (vs.\ 0.23 and 0.12 for Direct), so a fully open agent-and-model stack stays effective on a real production system.

\subsubsection{Case Study}
We trace one anonymized production incident across three consecutive days to show the self-evolution loop concretely. \textbf{(1)}~On day one, \texttt{/opsharness:setup} automatically samples the system's telemetry store and infers its data schema and layout, writing the system profile and $\mathcal{K}_1$ (the initial harness $h_0$). \textbf{(2)}~At 12:07 the next day, an upstream business service $\alpha$ alerts that entity-lookup queries keep failing, which triggers \texttt{/opsharness:diagnose}. Guided by $\mathcal{K}_0$, $\mathcal{K}_1$, and the profile, OpsHarness queries the telemetry and returns a ranked Top-3 root-cause list. An on-call SRE checks them in order: the first two are wrong, but the third is correct, an upstream database synchronization fault that leaves lookups repeatedly hitting empty records. The SRE confirms it with one line of feedback. \textbf{(3)}~Within the hour, at 13:00, \texttt{/opsharness:evolve}. mines the trajectory and distills the causal chain (DB-sync fault $\rightarrow$ ... $\rightarrow$ entity-query failure) and its supporting rules into a proposal. It passes the dual-gate verification and is committed as new $\mathcal{K}_2$/$\mathcal{K}_3$ knowledge ($h_i \rightarrow h_{i+1}$), and an SRE confirms the evolved items match the real failure pattern. \textbf{(4)}~On day three the same failure recurs, and OpsHarness now ranks the database-synchronization fault Top-1 on the first attempt, localizes the faulty component, and finishes in under 2 minutes, reusing a best practice mined automatically rather than written by hand.

\section{Discussion}
\label{sec:discuss}
\textbf{Performance of OpsHarness under system changes.} Such change is itself an argument for an evolving harness over a static one, since OpsHarness degrades gracefully rather than breaking. In the worst case a change invalidates the profile $\mathcal{K}_1$ and part of the mined knowledge, and OpsHarness falls back toward its cold-start state, which already matches or exceeds a bare agent (41.4\% vs.\ 36.1\% Final A@1) because $\mathcal{K}_0$ stays valid. Re-running \textit{setup} then rebuilds $\mathcal{K}_1$ with no code, and the harness re-accumulates from post-change diagnoses (Fig.~\ref{fig:rq2-curve}). Stale knowledge is retired rather than trusted, as the evolution action includes \textit{update} and \textit{delete} (Section~\ref{subsec:proposal}) that correct a rule once later trajectories contradict it.

\textbf{Failure case analysis.} We inspected the cases OpsHarness gets wrong and find two groups. Most are first encounters with a fault mode the harness has not seen before, so the agent reasons without the relevant experience and misses; these failures are progressively mitigated as self-evolution accumulates the corresponding best practices over later diagnoses (Fig.~\ref{fig:rq2-curve}). A second group is cases whose telemetry is too tangled for any reusable pattern to surface. We sampled these and find that over 75\% are also undiagnosable by human operators, which suggests they are close to unanswerable in the first place rather than a shortcoming specific to OpsHarness.

\section{Threats to Validity}
\label{sec:threats}
On \textit{internal} validity, LLM agents are non-deterministic, so we report aggregate accuracy over four backbones, two agent frameworks, and six sub-datasets rather than single runs.
On \textit{data leakage}, backbone memorization cannot explain our gains, since a bare agent on the same model reaches only 36.1\% Final A@1 (\textit{Direct}, Section~\ref{subsec:rq1}). We also keep training and test disjoint by a temporal split (Section~\ref{subsec:setup}), and a guard script blocks any access to ground-truth labels during diagnosis, so self-evolution never sees the answers it is later evaluated on.

\section{Conclusion}
\label{sec:conclusion}
With capable general agents, the leverage in LLM-based RCA moves from building an agent to building the harness around one. We present OpsHarness, the first self-evolving external RCA harness, pairing a data plane of layered knowledge and idea-card tools with a control plane. Its self-evolution mines best practices from correct and incorrect diagnoses and gates each change on a dual-gate verification. Across three benchmarks, four backbones, two agent frameworks, and an industrial deployment, OpsHarness achieves a 63.4\% improvement in average accuracy over the two bare agents, and a 4.02× improvement over baseline RCA agents on average.

\bibliographystyle{IEEEtran}
\bibliography{references}

\end{document}

%% file: tabs/exp_1.tex
\begin{table*}[t]
\centering
\caption{Overall effectiveness of 24 instances (four backbones $\times$ six frameworks) on the six sub-datasets of OpenRCA and RCAEval. 
\vspace{-0.10in}
\textnormal{
Specialized RCA agents are marked $\dagger$. Within each backbone block, \textbf{OpsHarness} (our method) is in \textbf{bold} and the best value in each column is \underline{underlined}.}}
\label{tab:rq1-main}
\setlength{\tabcolsep}{3pt}
\resizebox{\textwidth}{!}{%
\begin{tabular}{@{}l ccc ccc ccc ccc ccc ccc c@{}}
\toprule
\multirow{3}{*}{Framework}
 & \multicolumn{9}{c}{OpenRCA~\cite{openrca25}} & \multicolumn{9}{c}{RCAEval~\cite{rcaeval25}} & \multirow{3}{*}{\shortstack{Final\\A@1}} \\
\cmidrule(lr){2-10}\cmidrule(lr){11-19}
 & \multicolumn{3}{c}{Telecom} & \multicolumn{3}{c}{Bank} & \multicolumn{3}{c}{Market}
 & \multicolumn{3}{c}{Online Boutique} & \multicolumn{3}{c}{Sock Shop} & \multicolumn{3}{c}{Train Ticket} & \\
\cmidrule(lr){2-4}\cmidrule(lr){5-7}\cmidrule(lr){8-10}\cmidrule(lr){11-13}\cmidrule(lr){14-16}\cmidrule(lr){17-19}
 & A@1 & A@3 & Avg & A@1 & A@3 & Avg & A@1 & A@3 & Avg & A@1 & A@3 & Avg & A@1 & A@3 & Avg & A@1 & A@3 & Avg & \\
\midrule
\rowcolor{blue!10}\multicolumn{20}{c}{\textit{\textbf{GPT-5.5}}} \\
RCA-Agent$^{\dagger}$ & 36.4 & 45.5 & 55.0 & 35.7 & 64.3 & 65.0 & 23.2 & 53.6 & 55.0 & 0.0 & 11.1 & 46.0 & 5.6 & 5.6 & 29.0 & 0.0 & 0.0 & 24.0 & 16.8 \\
mABC$^{\dagger}$ & 9.1 & 18.2 & 14.0 & 10.7 & 21.4 & 35.0 & 3.1 & 9.8 & 32.0 & 16.7 & 33.3 & 76.0 & 11.1 & 22.2 & 61.0 & 5.6 & 11.1 & 61.0 & 9.4 \\
Codex (Direct) & 54.5 & 54.5 & 64.0 & 57.1 & 64.3 & 72.0 & 23.2 & 46.4 & 59.0 & 66.7 & 83.3 & 95.0 & 61.1 & 77.8 & 89.0 & 44.4 & 72.2 & 83.0 & 51.2 \\
Codex (ICL) & 27.3 & 36.4 & 50.0 & 46.4 & 53.6 & 65.0 & 20.0 & 40.0 & 49.0 & 55.6 & 77.8 & 93.0 & 61.1 & 61.1 & 85.0 & 61.1 & 88.9 & 94.0 & 45.3 \\
OpsHarness (no-evolve) & 54.5 & 54.5 & 64.0 & 46.4 & 57.1 & 62.0 & 26.8 & 43.3 & 58.0 & 66.7 & 72.2 & 91.0 & 72.2 & 72.2 & 91.0 & 50.0 & 77.8 & 89.0 & 52.8 \\
\textbf{OpsHarness} & \underline{\textbf{72.7}} & \underline{\textbf{72.7}} & \underline{\textbf{77.0}} & \underline{\textbf{64.2}} & \underline{\textbf{71.4}} & \underline{\textbf{78.0}} & \underline{\textbf{37.1}} & \underline{\textbf{66.5}} & \underline{\textbf{72.0}} & \underline{\textbf{72.2}} & \underline{\textbf{88.9}} & \underline{\textbf{96.0}} & \underline{\textbf{77.8}} & \underline{\textbf{88.9}} & \underline{\textbf{93.0}} & \underline{\textbf{72.2}} & \underline{\textbf{94.4}} & \underline{\textbf{96.0}} & \underline{\textbf{66.0}} \\
\midrule
\rowcolor{green!12}\multicolumn{20}{c}{\textit{\textbf{Claude Sonnet 4.6}}} \\
RCA-Agent$^{\dagger}$ & 9.1 & 18.2 & 32.0 & 53.6 & \underline{67.9} & 79.0 & 16.5 & 33.5 & 55.0 & 11.1 & 22.2 & 59.0 & 5.6 & 5.6 & 29.0 & 0.0 & 5.6 & 44.0 & 16.0 \\
mABC$^{\dagger}$ & 0.0 & 0.0 & 0.0 & 3.6 & 25.0 & 40.0 & 3.1 & 6.7 & 30.0 & 11.1 & 33.3 & 76.0 & 11.1 & 11.1 & 59.0 & 5.6 & 22.2 & 69.0 & 5.8 \\
Claude Code (Direct) & 36.4 & 45.5 & 55.0 & 24.2 & 36.3 & 61.0 & 17.0 & 26.3 & 38.0 & 38.9 & 61.1 & 85.0 & 27.8 & 44.4 & 76.0 & 16.7 & 22.2 & 76.0 & 26.8 \\
Claude Code (ICL) & 45.5 & 54.5 & 64.0 & 21.4 & 39.3 & 58.0 & 26.7 & 36.7 & 49.5 & 44.4 & 72.2 & 89.0 & 27.8 & 44.4 & 72.0 & 38.9 & 55.6 & 76.0 & 34.1 \\
OpsHarness (no-evolve) & 36.4 & \underline{63.6} & \underline{73.0} & 39.3 & 46.4 & 61.0 & 28.6 & 35.7 & 50.0 & 38.9 & 77.8 & 89.0 & 33.3 & 50.0 & 79.0 & 50.0 & 72.2 & 89.0 & 37.8 \\
\textbf{OpsHarness} & \underline{\textbf{63.6}} & \underline{\textbf{63.6}} & \underline{\textbf{73.0}} & \underline{\textbf{57.1}} & \textbf{63.7} & \underline{\textbf{81.9}} & \underline{\textbf{35.7}} & \underline{\textbf{64.2}} & \underline{\textbf{72.0}} & \underline{\textbf{61.1}} & \underline{\textbf{83.3}} & \underline{\textbf{95.0}} & \underline{\textbf{55.6}} & \underline{\textbf{77.8}} & \underline{\textbf{89.0}} & \underline{\textbf{61.1}} & \underline{\textbf{77.8}} & \underline{\textbf{93.0}} & \underline{\textbf{55.7}} \\
\midrule
\rowcolor{orange!13}\multicolumn{20}{c}{\textit{\textbf{GLM-5.2}}} \\
RCA-Agent$^{\dagger}$ & 36.4 & 54.5 & 59.0 & \underline{60.7} & \underline{64.3} & \underline{77.0} & 13.4 & 23.7 & 40.5 & 16.7 & 22.2 & 56.0 & 16.7 & 16.7 & 33.0 & 0.0 & 0.0 & 28.0 & 24.0 \\
mABC$^{\dagger}$ & 0.0 & 9.1 & 14.0 & 0.0 & 17.9 & 34.0 & 6.7 & 6.7 & 20.5 & 11.1 & 33.3 & 74.0 & 11.1 & 11.1 & 52.0 & 0.0 & 11.1 & 59.0 & 4.8 \\
Codex (Direct) & 54.5 & 63.6 & 68.0 & 57.1 & 57.1 & 67.0 & 26.8 & \underline{50.0} & \underline{65.0} & 38.9 & 77.8 & 93.0 & 44.4 & 72.2 & 89.0 & 55.6 & 72.2 & 87.0 & 46.2 \\
Codex (ICL) & 36.4 & 36.4 & 45.0 & 46.4 & 46.4 & 55.0 & 16.7 & 23.3 & 36.4 & 61.1 & 83.3 & \underline{95.0} & 50.0 & \underline{77.8} & \underline{91.0} & 50.0 & 77.8 & 87.0 & 43.4 \\
OpsHarness (no-evolve) & 54.5 & 63.6 & 80.0 & 57.1 & \underline{64.3} & 74.0 & 28.6 & 42.9 & 52.0 & 66.7 & \underline{88.9} & 89.0 & 44.4 & 55.6 & 74.0 & 55.6 & 88.9 & 94.0 & 51.2 \\
\textbf{OpsHarness} & \underline{\textbf{72.7}} & \underline{\textbf{81.8}} & \underline{\textbf{87.0}} & \textbf{57.1} & \underline{\textbf{64.3}} & \textbf{74.0} & \underline{\textbf{42.9}} & \underline{\textbf{50.0}} & \textbf{61.0} & \underline{\textbf{77.8}} & \underline{\textbf{88.9}} & \textbf{89.0} & \underline{\textbf{55.6}} & \textbf{66.7} & \textbf{76.0} & \underline{\textbf{88.9}} & \underline{\textbf{94.4}} & \underline{\textbf{98.0}} & \underline{\textbf{65.8}} \\
\midrule
\rowcolor{red!9}\multicolumn{20}{c}{\textit{\textbf{DeepSeek-V4}}} \\
RCA-Agent$^{\dagger}$ & \underline{45.5} & 45.5 & 50.0 & 28.6 & 35.7 & 47.0 & 9.8 & 9.8 & 26.0 & 5.6 & 5.6 & 35.0 & 0.0 & 5.6 & 26.0 & 0.0 & 0.0 & 22.0 & 14.9 \\
mABC$^{\dagger}$ & 0.0 & 0.0 & 5.0 & 3.6 & 14.3 & 32.0 & 0.0 & 3.1 & 16.0 & 11.1 & 33.3 & 72.0 & 0.0 & 5.6 & 43.0 & 0.0 & 0.0 & 48.0 & 2.5 \\
Codex (Direct) & 18.2 & 27.3 & 41.0 & 32.1 & 39.3 & 52.0 & 10.3 & 23.2 & 39.0 & 27.8 & 44.4 & 76.0 & 22.2 & 27.8 & 63.0 & 11.1 & 22.2 & 59.0 & 20.3 \\
Codex (ICL) & 36.4 & 36.4 & 41.0 & 28.6 & 39.3 & 50.0 & 20.0 & 30.0 & 42.8 & \underline{61.1} & 72.2 & \underline{91.0} & 22.2 & 33.3 & 61.0 & 16.7 & 38.9 & 67.0 & 30.8 \\
OpsHarness (no-evolve) & 27.2 & 45.5 & 46.0 & 35.7 & 46.4 & 61.0 & 21.4 & 28.6 & 50.0 & 44.4 & 53.3 & 76.0 & 27.8 & 50.0 & 80.0 & 33.3 & 66.7 & 72.0 & 31.6 \\
\textbf{OpsHarness} & \underline{\textbf{45.5}} & \underline{\textbf{63.6}} & \underline{\textbf{68.0}} & \underline{\textbf{46.4}} & \underline{\textbf{50.0}} & \underline{\textbf{64.0}} & \underline{\textbf{28.6}} & \underline{\textbf{42.9}} & \underline{\textbf{60.0}} & \textbf{53.3} & \underline{\textbf{73.3}} & \textbf{80.0} & \underline{\textbf{50.0}} & \underline{\textbf{72.2}} & \underline{\textbf{90.0}} & \underline{\textbf{66.7}} & \underline{\textbf{77.8}} & \underline{\textbf{89.0}} & \underline{\textbf{48.4}} \\
\bottomrule
\end{tabular}%
}
\vspace{-0.20in}
\end{table*}

%% file: references.bib
@inproceedings{fightincidents22,
  author    = {Supriyo Ghosh and Manish Shetty and Chetan Bansal and Suman Nath},
  title     = {How to Fight Production Incidents? An Empirical Study on a Large-Scale Cloud Service},
  booktitle = {Proc. 13th ACM Symp. Cloud Computing (SoCC)},
  pages     = {126--141},
  year      = {2022},
  doi       = {10.1145/3542929.3563482}
}

@article{faultdiagsurvey25,
  author  = {Shenglin Zhang and Sibo Xia and Wenzhao Fan and Binpeng Shi and Xiao Xiong and Zhenyu Zhong and Minghua Ma and Yongqian Sun and Dan Pei},
  title   = {Failure Diagnosis in Microservice Systems: A Comprehensive Survey and Analysis},
  journal = {ACM Trans. Softw. Eng. Methodol. (TOSEM)},
  year    = {2025},
  doi     = {10.1145/3715005},
  note    = {arXiv:2407.01710}
}

@inproceedings{microrca20,
  author    = {Li Wu and Johan Tordsson and Erik Elmroth and Odej Kao},
  title     = {{MicroRCA}: Root Cause Localization of Performance Issues in Microservices},
  booktitle = {Proc. IEEE/IFIP Network Operations and Management Symp. (NOMS)},
  pages     = {1--9},
  year      = {2020},
  doi       = {10.1109/NOMS47738.2020.9110353}
}

@inproceedings{sage21,
  author    = {Yu Gan and Mingyu Liang and Sundar Dev and David Lo and Christina Delimitrou},
  title     = {Sage: Practical and Scalable {ML}-Driven Performance Debugging in Microservices},
  booktitle = {Proc. 26th ACM Int. Conf. Architectural Support for Programming Languages and Operating Systems (ASPLOS)},
  pages     = {135--151},
  year      = {2021},
  doi       = {10.1145/3445814.3446700}
}

@inproceedings{eadro23,
  author    = {Cheryl Lee and Tianyi Yang and Zhuangbin Chen and Yuxin Su and Michael R. Lyu},
  title     = {{Eadro}: An End-to-End Troubleshooting Framework for Microservices on Multi-Source Data},
  booktitle = {Proc. 45th Int. Conf. Softw. Eng. (ICSE)},
  pages     = {1750--1762},
  year      = {2023},
  doi       = {10.1109/ICSE48619.2023.00150}
}

@article{baro24,
  author    = {Luan Pham and Huong Ha and Hongyu Zhang},
  title     = {{BARO}: Robust Root Cause Analysis for Microservices via Multivariate Bayesian Online Change Point Detection},
  journal   = {Proc. ACM Softw. Eng. (PACMSE), FSE},
  volume    = {1},
  number    = {FSE},
  pages     = {2214--2237},
  year      = {2024},
  doi       = {10.1145/3660805}
}

@inproceedings{rcacopilot24,
  author    = {Yinfang Chen and Huaibing Xie and Minghua Ma and Yu Kang and Xin Gao and Liu Shi and Yunjie Cao and Xuedong Gao and Hao Fan and Ming Wen and Jun Zeng and Supriyo Ghosh and Xuchao Zhang and Chaoyun Zhang and Qingwei Lin and Saravan Rajmohan and Dongmei Zhang and Tianyin Xu},
  title     = {Automatic Root Cause Analysis via Large Language Models for Cloud Incidents},
  booktitle = {Proc. 19th European Conf. Computer Systems (EuroSys)},
  pages     = {674--688},
  year      = {2024},
  doi       = {10.1145/3627703.3629553},
  note      = {arXiv:2305.15778}
}

@inproceedings{rcagent24,
  author    = {Zefan Wang and Zichuan Liu and Yingying Zhang and Aoxiao Zhong and Jihong Wang and Fengbin Yin and Lunting Fan and Lingfei Wu and Qingsong Wen},
  title     = {{RCAgent}: Cloud Root Cause Analysis by Autonomous Agents with Tool-Augmented Large Language Models},
  booktitle = {Proc. 33rd ACM Int. Conf. Information and Knowledge Management (CIKM)},
  pages     = {4966--4974},
  year      = {2024},
  doi       = {10.1145/3627673.3680016},
  note      = {arXiv:2310.16340}
}

@inproceedings{mabc24,
  author    = {Wei Zhang and Hongcheng Guo and Jian Yang and Yi Zhang and Chaoran Yan and Zhoujin Tian and Hangyuan Ji and Zhoujun Li and Tongliang Li and Tieqiao Zheng and Chao Chen and Yi Liang and Xu Shi and Liangfan Zheng and Bo Zhang},
  title     = {{mABC}: Multi-Agent Blockchain-Inspired Collaboration for Root Cause Analysis in Micro-Services Architecture},
  booktitle = {Findings of the Association for Computational Linguistics: EMNLP 2024},
  pages     = {4017--4033},
  year      = {2024},
  doi       = {10.18653/v1/2024.findings-emnlp.232}
}

@inproceedings{flowofaction25,
  author    = {Changhua Pei and Zexin Wang and Fengrui Liu and Zeyan Li and Yang Liu and Xiao He and Rong Kang and Tieying Zhang and Jianjun Chen and Jianhui Li and Gaogang Xie and Dan Pei},
  title     = {Flow-of-Action: {SOP} Enhanced {LLM}-Based Multi-Agent System for Root Cause Analysis},
  booktitle = {Companion Proc. ACM Web Conf. 2025 (WWW Companion)},
  year      = {2025},
  doi       = {10.1145/3701716.3715225},
  note      = {arXiv:2502.08224}
}

@misc{claudecode25,
  author       = {{Anthropic}},
  title        = {Claude Code},
  year         = {2025},
  howpublished = {\url{https://www.anthropic.com/claude-code}},
  note         = {Accessed: 2026-06}
}

@misc{openaicodex25,
  author       = {{OpenAI}},
  title        = {Introducing Codex},
  year         = {2025},
  howpublished = {\url{https://openai.com/index/introducing-codex/}},
  note         = {Accessed: 2026-06}
}

@article{sigelman2010dapper,
  title={Dapper, a large-scale distributed systems tracing infrastructure},
  author={Sigelman, Benjamin H and Barroso, Luiz Andre and Burrows, Mike and Stephenson, Pat and Plakal, Manoj and Beaver, Donald and Jaspan, Saul and Shanbhag, Chandan},
  year={2010}
}

@inproceedings{logad,
author = {Yu, Boxi and Yao, Jiayi and Fu, Qiuai and Zhong, Zhiqing and Xie, Haotian and Wu, Yaoliang and Ma, Yuchi and He, Pinjia},
title = {Deep Learning or Classical Machine Learning? An Empirical Study on Log-Based Anomaly Detection},
year = {2024},
isbn = {9798400702174},
publisher = {Association for Computing Machinery},
address = {New York, NY, USA},
url = {https://doi.org/10.1145/3597503.3623308},
doi = {10.1145/3597503.3623308},
booktitle = {Proceedings of the IEEE/ACM 46th International Conference on Software Engineering},
articleno = {35},
numpages = {13},
location = {Lisbon, Portugal},
series = {ICSE '24}
}

@inproceedings{Microscope,
  title={Microscope: Pinpoint Performance Issues with Causal Graphs in Micro-service Environments},
  author={Lin, Jinjin and Chen, Pengfei and Zheng, Zibin},
  booktitle={ICSOC 2018},
  pages={3--20},
  year={2018},
  organization={Springer}
}

@inproceedings{nezha,
author = {Yu, Guangba and Chen, Pengfei and Li, Yufeng and Chen, Hongyang and Li, Xiaoyun and Zheng, Zibin},
title = {Nezha: Interpretable Fine-Grained Root Causes Analysis for Microservices on Multi-modal Observability Data},
year = {2023},
isbn = {9798400703270},
publisher = {Association for Computing Machinery},
address = {New York, NY, USA},
url = {https://doi.org/10.1145/3611643.3616249},
doi = {10.1145/3611643.3616249},
booktitle = {Proceedings of the 31st ACM Joint European Software Engineering Conference and Symposium on the Foundations of Software Engineering},
pages = {553–565},
numpages = {13},
location = {San Francisco, CA, USA},
series = {ESEC/FSE 2023}
}

@inproceedings{lasrca,
author = {Han, Yongqi and Du, Qingfeng and Huang, Ying and Wu, Jiaqi and Tian, Fulong and He, Cheng},
title = {The Potential of One-Shot Failure Root Cause Analysis: Collaboration of the Large Language Model and Small Classifier},
year = {2024},
isbn = {9798400712487},
publisher = {Association for Computing Machinery},
address = {New York, NY, USA},
url = {https://doi.org/10.1145/3691620.3695475},
doi = {10.1145/3691620.3695475},
booktitle = {Proceedings of the 39th IEEE/ACM International Conference on Automated Software Engineering},
pages = {931–943},
numpages = {13},
location = {Sacramento, CA, USA},
series = {ASE '24}
}

@inproceedings{microdiag,
  TITLE = {{MicroDiag: Fine-grained Performance Diagnosis for Microservice Systems}},
  AUTHOR = {Wu, Li and Tordsson, Johan and Bogatinovski, Jasmin and Elmroth, Erik and Kao, Odej},
  URL = {https://inria.hal.science/hal-03155797},
  BOOKTITLE = {{ICSE21 Workshop on Cloud Intelligence}},
  ADDRESS = {Madrid, Spain},
  YEAR = {2021},
  MONTH = May,
  HAL_ID = {hal-03155797},
  HAL_VERSION = {v1},
}

@misc{openai2026gpt55,
  author       = {{OpenAI}},
  title        = {Introducing {GPT-5.5}},
  year         = {2026},
  howpublished = {\url{https://openai.com/index/introducing-gpt-5-5/}},
  note         = {Accessed: 2026-07-01}
}

@misc{anthropic2026sonnet46,
  author       = {{Anthropic}},
  title        = {Introducing {Claude Sonnet 4.6}},
  year         = {2026},
  howpublished = {\url{https://www.anthropic.com/news/claude-sonnet-4-6}},
  note         = {Accessed: 2026-07-01}
}

@misc{glm2026glm5,
  author        = {{GLM-5 Team}},
  title         = {{GLM-5}: From Vibe Coding to Agentic Engineering},
  year          = {2026},
  eprint        = {2602.15763},
  archivePrefix = {arXiv},
  primaryClass  = {cs.CL},
  url           = {https://arxiv.org/abs/2602.15763}
}

@misc{deepseek2026v4,
  author        = {{DeepSeek-AI}},
  title         = {{DeepSeek-V4}: Towards Highly Efficient Million-Token Context Intelligence},
  year          = {2026},
  eprint        = {2606.19348},
  archivePrefix = {arXiv},
  primaryClass  = {cs.CL},
  url           = {https://arxiv.org/abs/2606.19348}
}

@INPROCEEDINGS{cloudranger,
  author={Wang, Ping and Xu, Jingmin and Ma, Meng and Lin, Weilan and Pan, Disheng and Wang, Yuan and Chen, Pengfei},
  booktitle={2018 18th IEEE/ACM International Symposium on Cluster, Cloud and Grid Computing (CCGRID)}, 
  title={CloudRanger: Root Cause Identification for Cloud Native Systems}, 
  year={2018},
  volume={},
  number={},
  pages={492-502},
  doi={10.1109/CCGRID.2018.00076}}

@misc{icl,
      title={A Survey on In-context Learning}, 
      author={Qingxiu Dong and Lei Li and Damai Dai and Ce Zheng and Jingyuan Ma and Rui Li and Heming Xia and Jingjing Xu and Zhiyong Wu and Tianyu Liu and Baobao Chang and Xu Sun and Lei Li and Zhifang Sui},
      year={2024},
      eprint={2301.00234},
      archivePrefix={arXiv},
      primaryClass={cs.CL},
      url={https://arxiv.org/abs/2301.00234}, 
}

@inproceedings{openrca25,
  author    = {Junjielong Xu and Qinan Zhang and Zhiqing Zhong and Shilin He and Chaoyun Zhang and Qingwei Lin and Dan Pei and Pinjia He and Dongmei Zhang and Qi Zhang},
  title     = {{OpenRCA}: Can Large Language Models Locate the Root Cause of Software Failures?},
  booktitle = {Proc. Int. Conf. Learning Representations (ICLR)},
  year      = {2025},
  note      = {OpenReview: M4qNIzQYpd}
}

@inproceedings{roy-rcaagents24,
  author    = {Devjeet Roy and Xuchao Zhang and Rashi Bhave and Chetan Bansal and Pedro Las-Casas and Rodrigo Fonseca and Saravan Rajmohan},
  title     = {Exploring {LLM}-Based Agents for Root Cause Analysis},
  booktitle = {Companion Proc. ACM Int. Conf. Foundations of Softw. Eng. (FSE Companion)},
  pages     = {208--219},
  year      = {2024},
  doi       = {10.1145/3663529.3663841}
}

@misc{agentharness-anatomy26,
  author       = {Vivek Trivedy},
  title        = {The Anatomy of an Agent Harness},
  year         = {2026},
  howpublished = {LangChain Blog. \url{https://www.langchain.com/blog/the-anatomy-of-an-agent-harness}},
  note         = {Accessed: 2026-06}
}

@inproceedings{sweagent24,
  author    = {John Yang and Carlos E. Jimenez and Alexander Wettig and Kilian Lieret and Shunyu Yao and Karthik Narasimhan and Ofir Press},
  title     = {{SWE-agent}: Agent-Computer Interfaces Enable Automated Software Engineering},
  booktitle = {Advances in Neural Information Processing Systems (NeurIPS)},
  year      = {2024},
  note      = {arXiv:2405.15793}
}

@misc{agentskills25,
  author       = {{Anthropic}},
  title        = {Equipping Agents for the Real World with Agent Skills},
  year         = {2025},
  howpublished = {Anthropic Engineering. \url{https://www.anthropic.com/engineering/equipping-agents-for-the-real-world-with-agent-skills}},
  note         = {Accessed: 2026-06}
}

@misc{contexteng25,
  author       = {{Anthropic}},
  title        = {Effective Context Engineering for {AI} Agents},
  year         = {2025},
  howpublished = {Anthropic Engineering. \url{https://www.anthropic.com/engineering/effective-context-engineering-for-ai-agents}},
  note         = {Accessed: 2026-06}
}

@article{voyager24,
  author  = {Guanzhi Wang and Yuqi Xie and Yunfan Jiang and Ajay Mandlekar and Chaowei Xiao and Yuke Zhu and Linxi Fan and Anima Anandkumar},
  title   = {Voyager: An Open-Ended Embodied Agent with Large Language Models},
  journal = {Trans. Mach. Learn. Res. (TMLR)},
  year    = {2024},
  note    = {arXiv:2305.16291}
}

@inproceedings{expel24,
  author    = {Andrew Zhao and Daniel Huang and Quentin Xu and Matthieu Lin and Yong-Jin Liu and Gao Huang},
  title     = {{ExpeL}: {LLM} Agents Are Experiential Learners},
  booktitle = {Proc. AAAI Conf. Artificial Intelligence (AAAI)},
  volume    = {38},
  number    = {17},
  pages     = {19632--19642},
  year      = {2024},
  note      = {arXiv:2308.10144}
}

@misc{opencode25,
  author       = {{SST}},
  title        = {{OpenCode}: The Open Source {AI} Coding Agent},
  year         = {2025},
  howpublished = {\url{https://github.com/sst/opencode}},
  note         = {Accessed: 2026-06}
}

@inproceedings{autotsg22,
  author    = {Manish Shetty and Chetan Bansal and Sai Pramod Upadhyayula and Arjun Radhakrishna and Anurag Gupta},
  title     = {{AutoTSG}: Learning and Synthesis for Incident Troubleshooting},
  booktitle = {Proc. 30th ACM Joint European Softw. Eng. Conf. and Symp. Foundations of Softw. Eng. (ESEC/FSE), Industry Track},
  year      = {2022},
  doi       = {10.1145/3540250.3558958},
  note      = {arXiv:2205.13457}
}

@inproceedings{tsgrec20,
  author    = {Jiajun Jiang and Weihai Lu and Junjie Chen and Qingwei Lin and Pu Zhao and Yu Kang and Hongyu Zhang and Yingfei Xiong and Feng Gao and Zhangwei Xu and Yingnong Dang and Dongmei Zhang},
  title     = {How to Mitigate the Incident? An Effective Troubleshooting Guide Recommendation Technique for Online Service Systems},
  booktitle = {Proc. 28th ACM Joint Meeting European Softw. Eng. Conf. and Symp. Foundations of Softw. Eng. (ESEC/FSE), Industry Track},
  year      = {2020},
  doi       = {10.1145/3368089.3417054}
}

@inproceedings{saha-rckg22,
  author    = {Amrita Saha and Steven C. H. Hoi},
  title     = {Mining Root Cause Knowledge from Cloud Service Incident Investigations for {AIOps}},
  booktitle = {Proc. 44th Int. Conf. Softw. Eng.: Softw. Eng. in Practice (ICSE-SEIP)},
  year      = {2022},
  doi       = {10.1145/3510457.3513030},
  note      = {arXiv:2204.11598}
}

@inproceedings{ahmed-rcamitigation23,
  author    = {Toufique Ahmed and Supriyo Ghosh and Chetan Bansal and Thomas Zimmermann and Xuchao Zhang and Saravan Rajmohan},
  title     = {Recommending Root-Cause and Mitigation Steps for Cloud Incidents using Large Language Models},
  booktitle = {Proc. 45th Int. Conf. Softw. Eng. (ICSE)},
  year      = {2023},
  doi       = {10.1109/ICSE48619.2023.00149},
  note      = {arXiv:2301.03797}
}

@inproceedings{assesssummarize23,
  author    = {Pengxiang Jin and Shenglin Zhang and Minghua Ma and Haozhe Li and Yu Kang and Liqun Li and Yudong Liu and Bo Qiao and Chaoyun Zhang and Pu Zhao and Shilin He and Federica Sarro and Yingnong Dang and Saravan Rajmohan and Qingwei Lin and Dongmei Zhang},
  title     = {Assess and Summarize: Improve Outage Understanding with Large Language Models},
  booktitle = {Proc. 31st ACM Joint European Softw. Eng. Conf. and Symp. Foundations of Softw. Eng. (ESEC/FSE), Industry Track},
  year      = {2023},
  note      = {arXiv:2305.18084}
}

@inproceedings{iclrca-gpt424,
  author    = {Xuchao Zhang and Supriyo Ghosh and Chetan Bansal and Rujia Wang and Minghua Ma and Yu Kang and Saravan Rajmohan},
  title     = {Automated Root Causing of Cloud Incidents using In-Context Learning with {GPT-4}},
  booktitle = {Companion Proc. 32nd ACM Int. Conf. Foundations of Softw. Eng. (FSE Companion)},
  year      = {2024},
  doi       = {10.1145/3663529.3663846},
  note      = {arXiv:2401.13810}
}

@inproceedings{xpert24,
  author    = {Yuxuan Jiang and Chaoyun Zhang and Shilin He and Zhihao Yang and Minghua Ma and Si Qin and Yu Kang and Yingnong Dang and Saravan Rajmohan and Qingwei Lin and Dongmei Zhang},
  title     = {{Xpert}: Empowering Incident Management with Query Recommendations via Large Language Models},
  booktitle = {Proc. IEEE/ACM 46th Int. Conf. Softw. Eng. (ICSE)},
  year      = {2024},
  doi       = {10.1145/3597503.3639081},
  note      = {arXiv:2312.11988}
}

@misc{agentharness-survey26,
  author       = {Qianyu Meng and Yanan Wang and Liyi Chen and Wei Wu and Yihang Li and Wenyuan Jiang and Qimeng Wang and Chengqiang Lu and Yan Gao and Yi Wu and Yao Hu},
  title        = {Agent Harness for Large Language Model Agents: A Survey},
  year         = {2026},
  howpublished = {Preprints.org},
  doi          = {10.20944/preprints202604.0428.v1}
}

@misc{externalization26,
  author       = {Chenyu Zhou and Huacan Chai and Wenteng Chen and Zihan Guo and Rong Shan and Yuanyi Song and Tianyi Xu and Yingxuan Yang and Aofan Yu and Weiming Zhang and Congming Zheng and Jiachen Zhu and Zeyu Zheng and Zhuosheng Zhang and Xingyu Lou and Changwang Zhang and Zhihui Fu and Jun Wang and Weiwen Liu and Jianghao Lin and Weinan Zhang},
  title        = {Externalization in {LLM} Agents: A Unified Review of Memory, Skills, Protocols and Harness Engineering},
  year         = {2026},
  howpublished = {arXiv preprint arXiv:2604.08224}
}

@inproceedings{rcaeval25,
  author    = {Luan Pham and Hongyu Zhang and Huong Ha},
  title     = {{RCAEval}: A Benchmark for Root Cause Analysis of Microservice Systems with Telemetry Data},
  booktitle = {Companion Proc. ACM Web Conf. 2025 (WWW Companion)},
  year      = {2025},
  doi       = {10.1145/3701716.3715290},
  note      = {arXiv:2412.17015}
}

@article{coala24,
  author    = {Theodore R. Sumers and Shunyu Yao and Karthik Narasimhan and Thomas L. Griffiths},
  title     = {Cognitive Architectures for Language Agents},
  journal   = {Trans. Mach. Learn. Res. (TMLR)},
  year      = {2024},
  note      = {arXiv:2309.02427}
}

@misc{memgpt23,
  author       = {Charles Packer and Sarah Wooders and Kevin Lin and Vivian Fang and Shishir G. Patil and Ion Stoica and Joseph E. Gonzalez},
  title        = {{MemGPT}: Towards {LLMs} as Operating Systems},
  year         = {2023},
  howpublished = {arXiv preprint arXiv:2310.08560}
}

@misc{superpowers25,
  author       = {Jesse Vincent},
  title        = {{Superpowers}: An Agentic Skills Framework for Coding Agents},
  year         = {2025},
  howpublished = {\url{https://github.com/obra/superpowers}},
  note         = {Accessed: 2026-06}
}

@misc{ohmycodex25,
  author       = {Yeachan Heo},
  title        = {{oh-my-codex}: A Workflow Layer for the {OpenAI} Codex {CLI}},
  year         = {2025},
  howpublished = {\url{https://github.com/Yeachan-Heo/oh-my-codex}},
  note         = {Accessed: 2026-06}
}
